**WATERMARKING PROGRAM JAVA MENGGUNAKAN DUMMY METHODS DENGAN DYNAMICALLY OPAQUE PREDICATES**

**ZAENAL AKBAR**
**51 01 201 006**

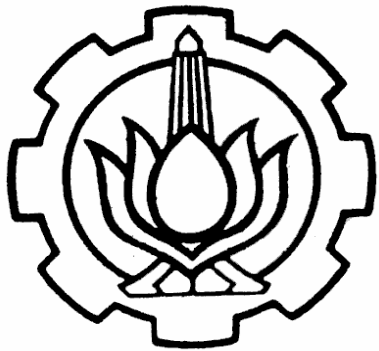

**PROGRAM PASCASARJANA**
**JURUSAN TEKNIK INFORMATIKA**
**FAKULTAS TEKNOLOGI INFORMASI**
**INSTITUT TEKNOLOGI SEPULUH NOPEMBER**
**SURABAYA**
**2004**

# T E S I S

## WATERMARKING PROGRAM JAVA MENGGUNAKAN DUMMY METHODS DENGAN DYNAMICALLY OPAQUE PREDICATES

Diajukan guna memenuhi salah satu persyaratan
memperoleh gelar Magister Komputer (M. Kom.)
pada
Program Pascasarjana
Jurusan Teknik Informatika
Fakultas Teknologi Informasi
Institut Teknologi Sepuluh Nopember
Surabaya

Mengetahui / Menyetujui
Dosen Pembimbing,

Prof. Ir. Handayani Tjandrasa, M.Sc., Ph.D.
NIP. 130 532 048

SURABAYA
2004

# WATERMARKING PROGRAM JAVA MENGGUNAKAN DUMMY METHODS DENGAN DYNAMICALLY OPAQUE PREDICATES

Tesis disusun guna memenuhi salah satu persyaratan
memperoleh gelar Magister Komputer (M. Kom.)
di
Institut Teknologi Sepuluh Nopember Surabaya

oleh :

**ZAENAL AKBAR**
**51 01 201 006**

Tanggal Ujian : 29 Januari 2004
Periode Wisuda : Maret 2004

**Disetujui oleh Tim Penguji Tesis :**   **Direktur Program Pascasarjana**

**Rully Sulaiman, S.Kom., M.Kom.**   **Prof. Ir. Happy Ratna S., M.Sc., Ph.D.**
**NIP. 132 085 802**   **NIP. 130 541 829**

**Febriliyan Samopa, S.Kom., M.Kom.**
**NIP. 132 206 858**

**Wiwik Anggraeni, S.Si., M.Kom.**
**NIP. 132 296 290**

# WATERMARKING PROGRAM JAVA MENGGUNAKAN DUMMY METHODS DENGAN DYNAMICALLY OPAQUE PREDICATES

**Zaenal Akbar**
Program Pascasarjana, Jurusan Teknik Informatika,
Fakultas Teknologi Informasi, Institut Teknologi Sepuluh Nopember,
Email : zaenal@programmer.net

## ABSTRAK

Pembajakan perangkat lunak dengan menggunakan, menggandakan dan menjual aplikasi secara tidak sah merupakan masalah utama yang menjadi perhatian para pembuat perangkat lunak. Selain itu, mereka juga khawatir aplikasi mereka akan di-*reverse engineer* dengan meng-ekstrak struktur data dan algoritma di dalamnya dan dipergunakan dalam aplikasi pesaing.

Salah satu metode untuk melindungi pembajakan perangkat lunak adalah *watermarking*, yaitu dengan menambahkan suatu pesan rahasia pada program yang ingin dilindungi. Metode ini tidak bertujuan untuk mencegah terjadinya pembajakan, tetapi sebagai tanda kepemilikan atas perangkat lunak termasuk jika terdapat struktur data dan algoritma yang digunakan pada aplikasi.

Program dalam bahasa Java didesain untuk dikompilasi dalam format *bytecode* dengan tujuan untuk dapat dieksekusi tanpa tergantung pada *platform* tertentu. Beberapa informasi pada kode sumber tidak berubah dalam *bytecode*, sehingga program lebih mudah di-dekompilasi kembali ke kode sumber dibandingkan dengan format *native-code*.

Tesis ini memberikan satu teknik untuk melindungi program Java, yaitu dengan menggunakan *dummy method* (method yang tidak akan pernah dieksekusi) yang dikembangkan dengan *dynamically opaque predicates* yang merupakan gabungan *opaque predicates* yang mempunyai nilai sama pada suatu *run* tertentu dan dapat mempunyai nilai lain saat *run* lainnya.

Setiap teknik *software watermarking* selalu berhubungan dengan *resilience*, *data rate*, *cost*, dan *stealth*. Untuk meng-evaluasi kualitas skema watermarking, terlebih dahulu harus diketahui jenis serangan yang mungkin. Teknik dalam tesis ini diharapkan tahan terhadap proses *translation* (kompilasi, dekompilasi, dan translasi biner), *optimization*, dan *obfuscation*.

Penambahan satu kode watermark menambah ukuran kode sumber sekitar 3.854 bytes untuk *dummy method* yang mampu menampung sampai 15 karakter dengan penambahan dua struktur data dinamis, dua buah *thread* serta penambahan dua *opaque predicates*. *Loading-time* aplikasi bertambah rata-rata 6108 milidetik

**Kata kunci** : *reverse engineer, software watermarking, dummy method, obfuscation, opaque predicates, dynamically opaque predicates.*

# WATERMARKING JAVA PROGRAMS USES DUMMY METHODS WITH DYNAMICALLY OPAQUE PREDICATES

**Zaenal Akbar**
Post Graduate Program, Informatics Engineering,
Faculty of Information Technology, Sepuluh Nopember Institute of Technology,
Email: zaenal@programmer.net

## ABSTRACT

Software piracy, the illegal using, copying, and resale of applications is a major concern for anyone develops software. Software developers also worry about their applications being *reverse engineered* by extracting data structures and algorithms from an application and incorporated into competitor's code.

A defense against software piracy is watermarking, a process that embeds a secret message in a cover software. Watermarking is a method that does not aim to stop piracy copying, but to prove ownership of the software and possibly even the data structures and algorithms used in the software.

The language Java was designed to be compiled into a platform independent *bytecode* format. Much of the information contained in the source code remains in the *bytecode*, which means that decompilation is easier than with traditional native codes.

In this thesis, we present a technique for watermarking Java programs by using a never-executed *dummy method*, that improved with *dynamically opaque predicates*, a family of *correlated* predicates which all evaluate to the same result in any given run, but in different runs they may evaluate to different results.

Any software watermarking technique will exhibit a trade-off between *resilience*, *data rate*, *cost*, and *stealth*. To evaluate the quality of a watermarking scheme we must also know how well it stands up to different types of *attacks*. Ideally, we would like our watermarks to survive *translation* (compilation, decompilation, and binary translation), *optimization*, and *obfuscation*.

Add a single watermark will increasing source code approximate 3.854 bytes with dummy method that cover up to 15 characters, two dynamic data structures, two threads and two opaque predicates. Application *loading-time* increase approximate 6108 milliseconds

# KATA PENGANTAR

Puji syukur penulis panjatkan kehadirat Allah SWT. atas segala limpahan berkah dan rahmat-Nya yang tiada henti-hentinya sejak penulis mulai menjadi mahasiswa pada Program Pascasarjana, Jurusan Teknik Informatika, Fakultas Teknologi Informasi, Institut Teknologi Sepuluh Nopember Surabaya, sampai tesis ini dapat terselesaikan. Semoga berkah dan rahmat-Nya terus melimpahi kita semua dalam menapaki hari-hari ke depan. Amin.

Pada kesempatan ini, penulis mengucapkan terima kasih yang sebesar-besarnya kepada :

1. **Prof. Ir. Muhammad Nuh, DEA., Ph.D.**, selaku Rektor Institut Teknologi Sepuluh Nopember Surabaya.

2. **Prof. Ir. Happy Ratna S., M.Sc., Ph.D.**, selaku Direktur Program Pascasarjana, Institut Teknologi Sepuluh Nopember Surabaya.

3. **Prof. Ir. Arif Djunaidy, M.Sc., Ph.D.**, selaku Dekan Fakultas Teknologi Informasi, Institut Teknologi Sepuluh Nopember Surabaya.

4. **Prof. Ir. Handayani Tjandrasa, M.Sc., Ph.D.**, selaku Dosen Pembimbing penulis, atas segala bantuan, arahan dan kesabaran hati yang diberikan sampai tesis ini dapat terselesaikan.

5. **Ir. FX. Arunanto, M.Sc.**, selaku Koordinator Program Pascasarjana, Program Studi Teknik Informatika, Institut Teknologi Sepuluh Nopember Surabaya.

6. **Ir. Esther Hanaya, M.Sc.**, selaku Dosen Wali penulis.

# DAFTAR ISI

# DAFTAR GAMBAR

# DAFTAR GRAFIK

# BAB I

# PENDAHULUAN

## 1.1 Latar Belakang

Banyaknya pembajakan perangkat lunak dengan menggunakan, menggandakan dan menjual (sebagian maupun keseluruhan) secara tidak sah merupakan masalah yang menjadi perhatian utama para pembuat dan penjual perangkat lunak [COL-2000]. Hal ini semakin diperparah dengan terus berkembangnya teknologi informasi, yang semakin mempermudah hal tersebut dilakukan.

Selain itu, banyak pembuat perangkat lunak juga khawatir jika aplikasi mereka akan di-*reverse engineer* [COL-1999, LOW-1998, LOW-1998A] yaitu mengambil algoritma dan struktur dalam aplikasi orang lain untuk digunakan dalam aplikasi sendiri. Hal ini dimungkinkan dengan menggunakan *disassembler* atau *decompiler* [CIF-1995] untuk merubah aplikasi menjadi kode sumber, yang selanjutnya struktur data maupun algoritma aplikasi akan dapat dianalisa.

Jika algoritma atau struktur data aplikasi diketahui oleh pihak lain dan digunakan pada aplikasi-nya sendiri, maka akan membuat biaya pembuatan aplikasi pihak lain tersebut menjadi lebih sedikit. Dengan demikian lama-kelamaan persaingan akan menjadi tidak sehat dan dapat membuat pembuat perangkat lunak yang algoritma dan struktur data-nya diambil pihak lain itu menjadi kalah bersaing bahkan tersingkir.

## 1.2 Rumusan Masalah

Bahasa Java didesain untuk dikompilasi dalam format *bytecode* yang *platform independent*. Beberapa informasi pada kode sumber masih tetap bentuknya dalam *bytecode* [PRO-1997], sehingga lebih mudah untuk dianalisa [DAH-1999]. Slogan "*write once, run anywhere*" adalah keunggulan sekaligus kekurangan bahasa Java.

Berbagai penelitian telah dilakukan untuk memperoleh teknik yang terbaik untuk perlindungan perangkat lunak, salah satu diantaranya adalah software watermarking. Teknik watermarking banyak diterapkan untuk perlindungan citra image, audio, atau video, sedangkan untuk perlindungan perangkat lunak masih kurang [COL-1999].

## 1.3 Tujuan Penelitian

Penelitian ini dimaksudkan untuk menemukan suatu teknik terbaik dalam melindungi program Java (dalam hal ini *class file*) dengan menggunakan *dummy method watermarking* [MON-2000] yang digabungkan dengan *opaque predicates obfuscation* [ARB-2002] dan dikembangkan dengan *dynamically opaque predicates*.

Hal utama yang ingin diteliti adalah mengenai penggunaan *opaque predicate* yang bersifat dinamis (*dynamically opaque predicates*) dengan tujuan untuk semakin mengaburkan keberadaan *dummy method* (yang merupakan lokasi *watermark* dalam program), sehingga semakin menyulitkan pihak-pihak lain untuk mendeteksi keberadaan *watermark* atau bahkan tidak dapat mengetahui-nya.

**1.4 Ruang Lingkup**

Pembahasan dalam tesis ini dibagi dalam dua bagian besar dalam hubungannya dengan software watermarking, yaitu :

*Dummy Method Watermarking* : Bagian ini mencakup proses memasukkan kode watermark dalam program. Kode ini diikutkan dalam suatu method kosong (*dummy*) yang sengaja disiapkan untuk hal tersebut.

*Dynamically Opaque Predicates* : Bagian ini akan membahas penerapan *opaque predicate* yang bersifat dinamis dalam rangka mengaburkan letak *dummy method* dalam suatu program. Semakin kabur letak *dummy method* (yang merupakan letak kode watermark) berarti aplikasi akan semakin aman karena tanda kepemilikan akan selalu ada dalam program tersebut.

Penelitian dalam bidang software watermarking adalah demikian luas, sehingga beberapa batasan awal yang dapat disebutkan adalah :

1. Program yang digunakan sebagai obyek untuk di-watermark adalah yang menggunakan bahasa Java. Hal ini dikarenakan program dalam bahasa Java lebih mudah untuk di-*reverse engineer* dibandingkan dengan program yang ber-format *binary-code* . Namun tidak menutup kemungkinan bahwa metode yang diteliti ini juga dapat diterapkan pada bahasa pemrograman lainnya.
2. Opaque predicate yang bersifat dinamis dapat mempunyai bentukan yang sangat banyak yang tergantung pada cara pengelompokan opaque-opaque predicate tersebut. Penelitian ini dibatasi pada penggunaan kelompok yang anggotanya tetap.

Tingkat keberhasilan penelitian ini sangat tergantung pada tingkat resistansi hasil penelitian terhadap serangan (*attack*) yang diberikan. Attack dapat dilakukan secara manual maupun otomatis dengan menggunakan tool-tool tertentu. Untuk attack yang dilakukan secara manual akan sulit dalam mengukur tingkat resistansi hasil penelitian (tergantung pengetahuan attacker—yang berbeda-beda), sehingga attack yang akan dilakukan dibatasi pada penggunaan tool-tool otomatis yang sudah ada.

Secara keseluruhan tesis ini akan menjelaskan secara teknis suatu metode perlindungan perangkat lunak yang dapat dilakukan oleh pemilik yang sah. Namun demikian, hal tersebut harus dibarengi dengan peraturan pemerintah yang tentunya harus memberi perlindungan atau pengakuan pada teknik ini, misalnya saja dengan pemberian jaminan bahwa kode rahasia (watermark) dalam program dapat dijadikan bukti di pengadilan.

**1.5 Metodologi Penelitian**

1. Studi kepustakaan

    Studi kepustakaan dilakukan untuk memperdalam pengetahuan mengenai dummy method watermarking, opaque predicates terutama untuk pembentukan opaque predicates yang dinamis.

2. Desain sistem

    a. Dummy Method Watermarking :

    - Desain algoritma pengkodean watermark dalam program, termasuk kemungkinan penggunaan kode berbentuk dinamis.

    - Desain algoritma pen-dekodean watermark.

- Desain penggunaan kunci (key) untuk pengkodean dan pen-dekodean watermark untuk lebih mempersulit pendeteksian watermark.
- Desain sistem pengkodean dan pen-dekodean watermark secara otomatis.

b. Dynamically Opaque Predicates :
- Desain algoritma pengelompokan opaque predicates sehingga nilai dari setiap kelompok sesuai dengan yang diharapkan.

3. Implementasi sistem

Sistem diimplementasikan dalam bahasa pemrograman sesuai dengan desain yang sudah ada.

4. Pengujian dan analisa sistem

Tahap pertama, yang dilakukan secara off-line :

- Memasukkan satu atau beberapa method dummy pada kode sumber.
- Membuat opaque-opaque predicate pada bagian-bagian tertentu (secara acak) dari kode sumber.
- Mengelompokkan opaque predicates tersebut dan menghubungkan method dummy dengan method lain sehingga menjadi satu kesatuan.
- Melakukan kompilasi program sehingga diperoleh class-file.

Tahap kedua, yang dilakukan secara off-line maupun on-line:

- Kode watermark di-injeksi dalam method dummy pada class-file
- Memberikan serangan (*obfuscator attack*, *decompile-compile attack*, dan lain-lain) dengan tool-tool yang sudah ada.
- Serangan dapat juga berbentuk dynamic analisis, misalnya dengan memonitor heap, register atau memory mesin ketika program dijalankan.

- Men-dekode watermark dari program yang telah diberi serangan.

Hasil pengujian kemudian dianalisa, dan apabila dianggap belum memadai dapat kembali ke langkah 2.

5. Dokumentasi

Semua proses dan hasil yang diperoleh didokumentasikan, juga merupakan bahan analisis kembali.

## 1.6 Sistematika Pembahasan

Sistematika pembahasan dalam tesis ini dapat diuraikan sebagai berikut :

**Bab I. Pendahuluan.**

Bagian ini membahas latar belakang permasalahan yang diteliti, tujuan dan ruang lingkup penelitian serta metodologi yang digunakan dalam penelitian.

**Bab II. Landasan Teori.**

Bagian ini berisi teori-teori hasil studi kepustakaan mengenai software watermarking secara umum dan khususnya mengenai dummy method watermarking serta opaque predicates.

**Bab III. Desain dan Implementasi Sistem.**

Bagian ini membahas perancangan sistem yang diterapkan untuk menyelesaikan permasalahan disertai dengan implementasi-nya.

**Bab IV. Pengujian dan Analisa Sistem**

Sistem yang telah dirancang dan dibangun selanjutnya di-ujicobakan dengan menggunakan parameter-parameter umum yang digunakan dalam software watermarking. Pengujian, hasil pengujian serta analisa hasil yang diperoleh dibahas pada bagian ini.

**Bab V. Kesimpulan dan Saran.**

Dari hasil analisa sebelumnya, dapat ditarik kesimpulan-kesimpulan serta saran-saran dalam rangka pengembangan lebih lanjut. Bagian ini berisikan hal tersebut.

# BAB II

# LANDASAN TEORI

## 2.1 Perlindungan Perangkat Lunak

Perangkat lunak merupakan hasil karya manusia yang dalam proses pembuatannya membutuhkan berbagai sumber daya, seperti tenaga, waktu ataupun uang. Sama dengan hasil karya lainnya, perangkat lunak juga perlu dilindungi sebagai salah satu bentuk pengakuan atas hasil karya tersebut.

Secara umum perlindungan perangkat lunak dilatar belakangi oleh dua jenis serangan yang mungkin terjadi [COL-2000]:

1. Serangan dari *malicious client programs*.

   Serangan jenis ini dilakukan oleh suatu program yang melakukan tindakan tertentu pada host yang diserang seperti merusak data (gambar 2.1 (a)). Contoh serangan jenis ini adalah virus-komputer.

   Untuk menghadapi serangan jenis ini, biasanya host membatasi aksi yang bisa dilakukan oleh program, misalnya tidak boleh ada penulisan pada file system. Salah satu teknik perlindungan dari serangan jenis ini adalah *Software Fault Isolation* [WAH-1993].

2. Serangan dari *malicious hosts*.

   Pada serangan jenis ini, program mendapat ancaman dari host dimana program di-download dan atau dijalankan (gambar 2.1 (b)). Serangan ini umumnya menyebabkan terjadinya pelanggaran hak intelektual. Program mungkin berisi materi-materi penting yang jika diketahui akan menimbulkan kerugian finansial pada pembuatnya.

(a) Serangan dari *malicious client program*

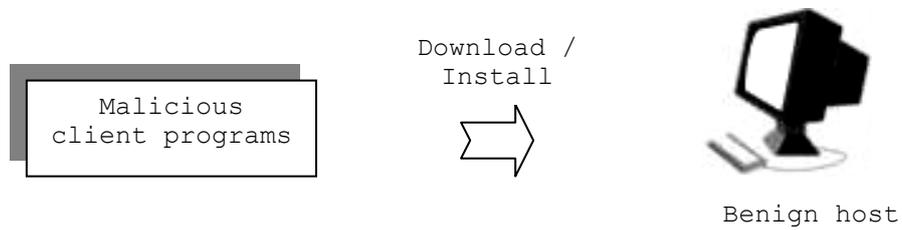

(b) Serangan dari *malicious host*

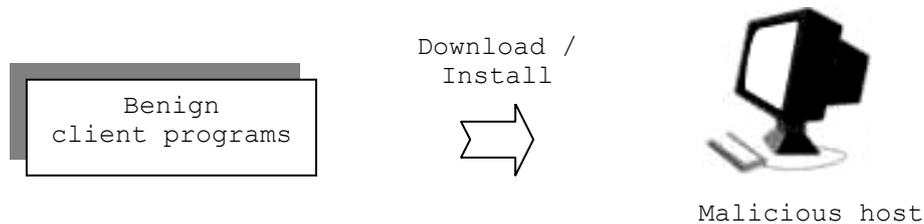

*Gambar 2.1  Jenis serangan terhadap perangkat lunak*

### 2.1.1  Perlindungan Hak Intelektual dalam Perangkat Lunak [COL-2000]

Pelanggaran hak intelektual dimungkinkan pada malicious host karena program ter-download dan atau ter-install pada host. Dengan kata lain, program secara fisik berada dalam host.

Ada tiga jenis serangan pada malicious host :

1. Pembajakan perangkat lunak.

    Pembajakan perangkat lunak adalah menggunakan, memperbanyak dan menjual kembali secara tidak sah. Nilai pembanyakan sampai mencapai 15 milyar dolar per-tahun [COL-2000]. Nilai ini tentu saja dapat berubah setiap waktu. Hal ini menyebabkan pembajakan perangkat lunak mendapat perhatian besar dari mereka yang membuat dan menjual perangkat lunak.

Pada gambar 2.2 (a), *B* membuat salinan aplikasi yang secara sah diperoleh dari *A*, kemudian menjual-nya secara tidak sah kepada para pembeli.

2. *Reverse engineering.*

    Selain itu, banyak pembuat perangkat lunak juga khawatir aplikasi mereka akan di-*reverse engineer* [COL-1999, LOW-1998, LOW-1998A] yaitu dengan menggunakan *disassembler* atau *decompiler* [CIF-1995] untuk merubah program kembali menjadi kode sumber, kemudian struktur data maupun algoritmanya akan dapat dianalisa, yang selanjutnya dapat saja digunakan (sebagian atau seluruh aplikasi ) pada aplikasi buatannya sendiri.

    Pada gambar 2.2 (b), *B* men-dekompilasi dan me-reverse engineer aplikasi yang diperoleh dari *A*, yang selanjutnya mengambil satu atau beberapa materi di dalamnya untuk digunakan pada aplikasi sendiri.

3. *Tampering.*

    Banyak aplikasi (misal untuk e-commerce) berisi kode, kunci atau informasi rahasia yang ter-enkripsi. Seseorang yang mampu meng-ekstrak atau memodifikasi rahasia tersebut akan menyebabkan kerugian finansial pada pemilik yang sah.

    Pada gambar 2.2 (c), *B* menerima "digital container" dari *A* yang berisi beberapa media digital termasuk kode yang akan men-transfer sejumlah uang kepada account *A* jika media tersebut dimainkan. *B* dapat melakukan perubahan jumlah yang harus dibayarkan atau bahkan meng-ekstrak media digital-nya saja. Untuk kasus terakhir, *B* dapat terus menikmati media secara gratis bahkan menjual pada pihak lain.

(a) *Pembajakan perangkat lunak*, B membuat salinan program P milik A dan menjualnya secara tidak sah.

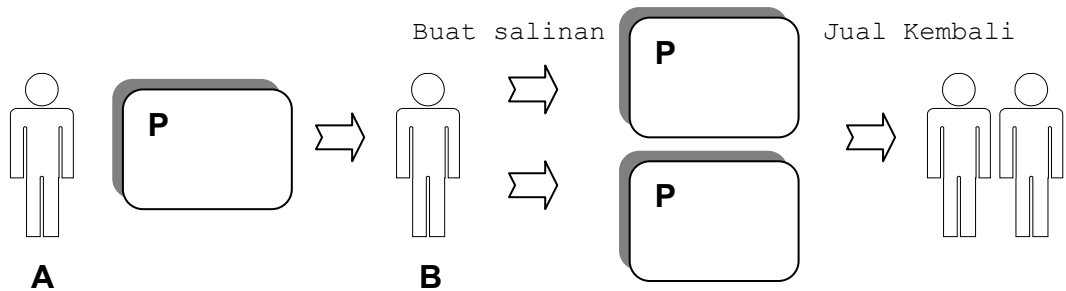

(b) *Reverse engineering*, B meng-ekstrak modul M dari program P milik A dan menggunakan pada program Q miliknya.

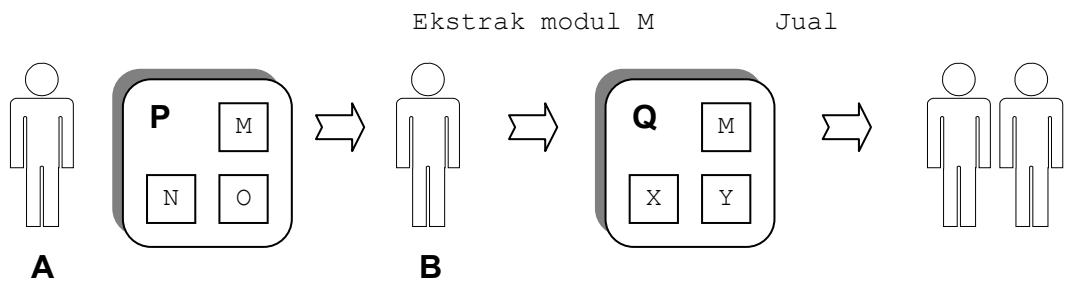

(c) *Tampering*, B meng-ekstrak media dari container milik A atau merubah nilai yang akan dibayarkan.

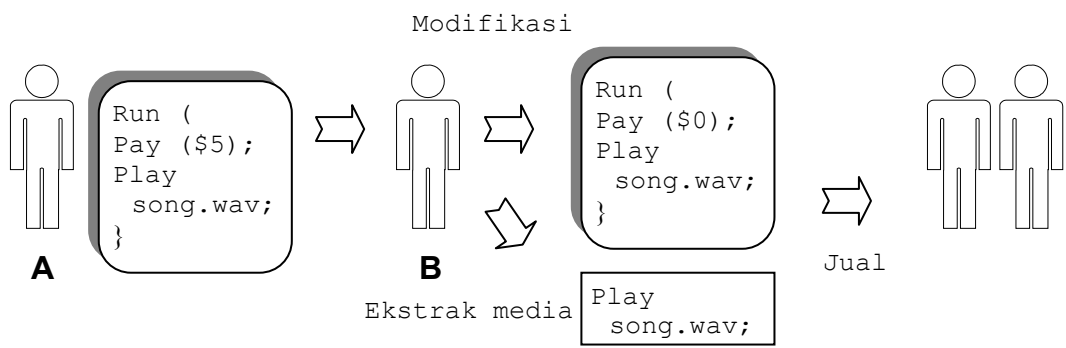

*Gambar 2.2 Serangan terhadap hak intelektual dalam perangkat lunak*

Banyak penelitian dilakukan untuk menghadapi serangan-serangan tersebut terutama serangan pada *malicious host*. Hal ini disebabkan perlindungan untuk serangan *malicious host* sangat terbatas. Sekali program berada pada mesin host, segala cara dapat dilakukan untuk meng-ekstrak hal-hal penting didalamnya.

Penelitian Collberg dan Thomborson [COL-2000] menghasilkan beberapa cara untuk menghadapi serangan *malicious host* tersebut, yaitu :

1. *Software watermarking* untuk pembajakan.

   Software watermarking adalah proses pemberian kode rahasia dalam aplikasi yang menjadi tanda kepemilikan atas aplikasi tersebut.

   Pada gambar 2.3 (a), *A* memberikan kode *W* dalam program *P* sebagai watermark dengan menggunakan kunci *K*. Jika *B* melakukan pencurian program *P*, maka *C* dapat men-dekoding *W* menggunakan key *K* untuk membuktikan bahwa *P* adalah benar milik *A*.

2. *Tamper-proofing* untuk *Tampering*.

   Tamper-proofing adalah proses pemberian kode yang akan membuat aplikasi tidak menjadi tidak berfungsi apabila mengalami perubahan.

   Pada gambar 2.3 (b), *A* memberikan kode tamper-proofing *S* dalam *P*. Karena *B* melakukan perubahan program *P*, maka *S* akan membuat *P* menjadi rusak.

3. *Obfuscation* untuk reverse-engineering

   Obfuscation adalah proses men-transformasi suatu aplikasi ke bentuk lain yang mempunyai sifat sama dengan aslinya, akan tetapi lebih sulit untuk dimengerti.

   Pada gambar 2.3 (c), *A* memberikan obfuscation transformation $T_1$, $T_2$, $T_3$ yang akan menyulitkan *B* melakukan reverse-engineering.

(a) *Software watermarking*, A memberikan kode watermark W dengan kunci K, C dapat menggunakan K untuk men-dekoding W.

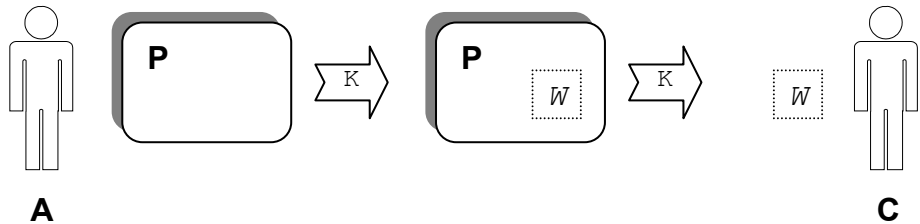

(b) *Tamper-proofing*, (1)A memberikan kode tamper-proofing S dalam P (2)B melakukan modifikasi sehingga P menjadi rusak.

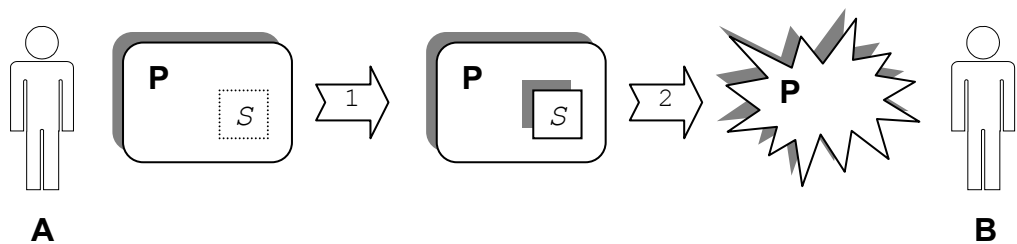

(c) *Obfuscation*, A melakukan transformasi obfuscation $T_1$, $T_2$, $T_3$ sehingga B kesulitan me-reverse engineer P.

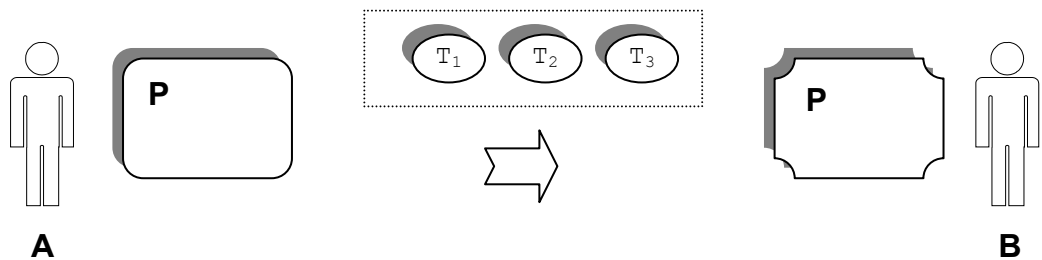

*Gambar 2.3  Perlindungan dari serangan malicious host*

## 2.1.2  Perlindungan Aplikasi dalam Bahasa Java

Program dalam bahasa Java dapat berbentuk applet untuk aplikasi web pada internet maupun berbentuk aplikasi mandiri seperti aplikasi dari bahasa lainnya. Untuk menjalankan kedua jenis aplikasi bahasa Java ini, program yang berbentuk *class file* harus berada pada mesin host. Hal ini menyebabkan serangan malicious host sangat mungkin terjadi. Untuk itu penelitian ini difokuskan untuk mengatasi serangan yang termasuk dalam jenis malicious host attack ini.

Dari ketiga jenis perlindungan untuk malicious host attack yang dijelaskan pada bagian sebelumnya, kesemuanya dapat diterapkan untuk melindungi aplikasi dalam bahasa Java, baik secara individual maupun gabungannya. Secara umum apabila ketiga teknik tersebut digunakan bersama-sama, maka aplikasi akan semakin aman.

Penelitian ini mencoba melakukan penggabungan dua teknik yaitu software watermarking dan obfuscation, sementara tamper-proofing tidak termasuk bagian penelitian. Namun, sekilas teknik tamper-proofing hampir sama dengan software watermarking yaitu sama-sama menambahkan kode rahasia dalam aplikasi. Kalau pada software watermarking, kode tersebut menjadi tanda maka pada tamper-proofing kode tersebut merupakan bagian aplikasi yang akan memeriksa apakah telah terjadi perubahan pada aplikasi.

## 2.2 Bahasa Pemrograman Java.

Bahasa Java merupakan bahasa pemrograman yang berorientasi obyek. Platform Java dibangun untuk mengatasi masalah dalam membuat perangkat lunak yang ditujukan bagi peralatan pengguna yang menggunakan jaringan. Ini didesain untuk dapat mendukung berbagai macam aristektur dan memungkinkan pengiriman komponen perangkat lunak secara aman. Untuk memenuhi hal tersebut, kode yang telah dikompilasi harus tetap aman dalam pengiriman pada jaringan, dapat dioperasikan pada sembarang client dan aman untuk dijalankan.

## 2.2.1 Java Virtual Machine [LIN-1999]

*Java virtual machine* adalah mesin perhitungan abstrak yang seperti mesin perhitungan sebenarnya, juga mempunyai sejumlah instruksi dan memanipulasi sejumlah lokasi memori pada saat run time. Virtual machine ini tidak mengetahui apapun tentang bahasa pemrograman Java, hanya sedikit format binary, yang disebut format *class file*. Suatu class file berisi instruksi-instruksi Java virtual machine (*bytecodes*) dan suatu *symbol table*.

Kode sumber bahasa Java dikompilasi oleh Java compiler menjadi java class file. File ini kemudian dimuat oleh *Java Loader* (dapat secara local maupun melalui jaringan). *Java class library* yang dibutuhkan juga dimuatkan pada tahap ini. Sebelum class file ini dieksekusi, terlebih dahulu di-cek oleh *Java verifier*. Jika tidak ditemukan adanya kesalahan, Java virtual machine akan meng-eksekusi class file tersebut (gambar 2.4).

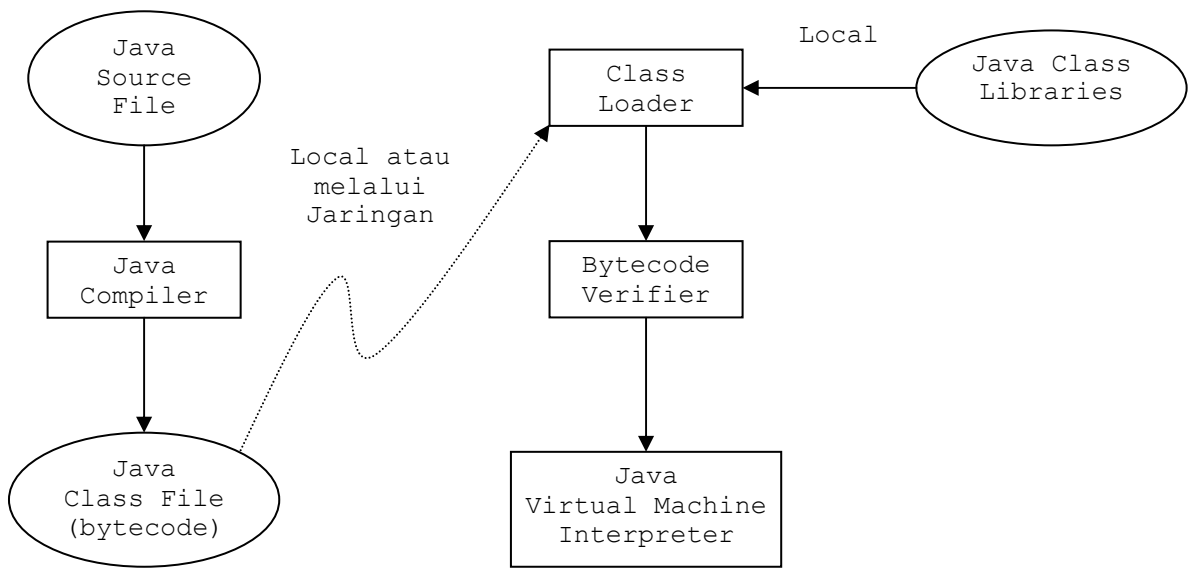

*Gambar 2.4 Proses kompilasi kode sumber bahasa java dan eksekusi Java class file*

## 2.2.2 Format Java Class File [DAH-1999]

Setiap class file berisi defenisi satu class atau interface, yang terdiri atas *stream-stream* 8-bit per-byte-nya. Untuk pemakaian 16-bit, 32-bit, dan 64-bit di bentuk dengan membaca stream sebanyak dua, empat dan delapan kali. Item data yang lebih dari satu byte disimpan dalam susunan *big-endian*, dimana byte yang lebih besar datang lebih dahulu. Struktur format Java class file dapat dilihat secara lengkap pada lampiran I.

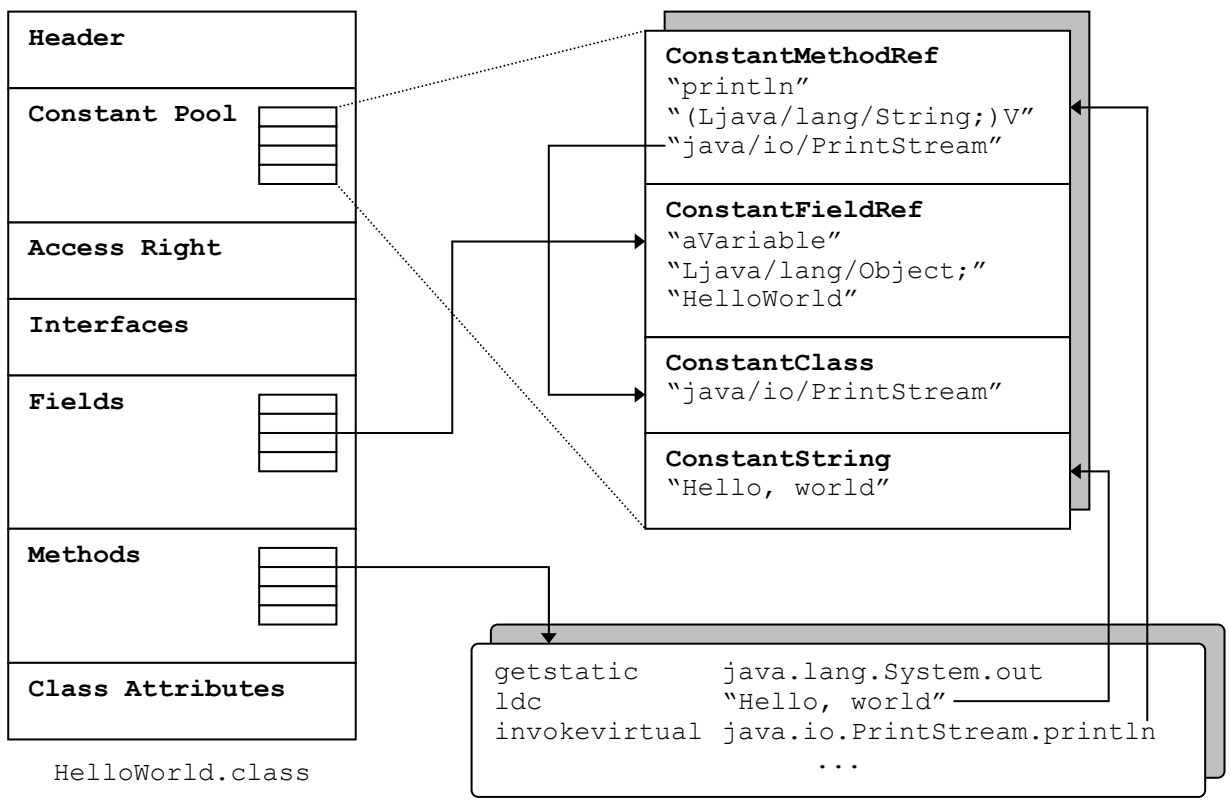

*Gambar 2.5 Format Java Class File untuk program HelloWorld*

## 2.2.3 Serangan Terhadap Program Java

Dengan tujuan untuk portabilitas yaitu dapat dieksekusi dimana saja tanpa tergantung pada suatu platform tertentu, program Java didesain untuk dikompilasi dalam format bytecode. Beberapa informasi pada kode sumber masih tetap bentuknya dalam bytecode [PRO-1997], sehingga lebih mudah untuk dianalisa [DAH-1999].

Dengan menggunakan *class editor* seperti *CFParse* dari *IBM AlphaWorks*, *BCEL* , *BLOAT* dapat diketahui cara penggunaan class tersebut tanpa bertanya pada pembuatnya bahkan dapat dilakukan modifikasi. Kode sumber dari class file juga dapat di-*generate* dengan menggunakan Java decompiler, seperti *Jad*, *Jode, Mocha*.

## 2.3 Software Watermarking

Watermarking menempelkan kode rahasia dalam perangkat lunak yang ingin dilindungi. Pada *media watermarking*, kode tersebut biasanya merupakan pesan *copyright*. Watermarking dapat mencegah pencurian, dan apabila terjadi pencurian memungkinkan pemilik untuk meng-klaimnya.

*Fingerprinting* mirip dengan watermarking, perbedaan hanya pada kode rahasia yang dimasukkan dalam program yang tidak sama untuk setiap distribusinya. Ini memungkinkan tidak hanya untuk pendeteksian terjadinya pencurian, tetapi juga mengetahui pengguna yang melanggar *copyright*

Permasalahan software watermarking dapat dijelaskan sebagai berikut [COL-1999] :

Tambahkan struktur *W* (watermark) dalam program *P* sedemikian sehingga :

- *W* dapat dilokalisir dan di-ekstrak dari *P* (program) dimanapun *W* berada dalam *P*, termasuk jika *P* sudah mengalami proses transformasi (*resilient*),
-  *W* cukup besar (*high data rate*),
- Penambahan *W* dalam *P* tidak merubah performance *P* (c*heap*),

- Penambahan *W* dalam *P* tidak merubah statistical properties *P* (s*tealthy*),
- *W* mempunyai property matematis sehingga memungkinkan pemilik untuk ber-argumentasi bahwa keberadaan *W* dalam *P* merupakan hasil proses matematik.

### 2.3.1 Aplikasi Software Watermarking [NAG-2002]

Penerapan software watermarking dapat mempunyai tujuan yang berbeda, tergantung pada hal utama yang ingin diberikan perlindungan. Walaupun berbeda dalam penerapan-nya, konsep atau metode yang digunakan saja sama. Terdapat tiga bagian penting dalam penerapan watermarking ini :

(1) *Author*

Watermarking pada umumnya diterapkan untuk meng-identifikasi pembuat (author) dari suatu perangkat lunak dan melindungi hak intelektual-nya. Pembuat dapat terdiri dari satu orang ataupun beberapa orang yang mana dapat saja mempunyai kode watermark yang berbeda-beda. Kode ini disebut sebagai *Authorship Marks*.

**Authorship Mark (AM)** adalah watermark yang ditempelkan dalam perangkat lunak untuk meng-identifikasi-kan pembuatnya (author).

Authorship Marks dapat meng-identifikasi pembuat tunggal, yang disebut *Single Authorship Marks*. Hal lainnya, juga dapat meng-identifikasi sejumlah pembuat, yang disebuat *Multiple Authorship Marks*.

(2) *Distributor*

Watermarking juga dapat digunakan untuk mengetahui jalur distribusi dan selanjutnya dapat meng-identifikasi legal atau ilegal salinan suatu perangkat lunak. Watermark jenis ini disebut sebagai *Fingerprint Mark* atau suatu *Fingerprint*.

**Fingerprint Mark (FM)** adalah watermark yang menempelkan informasi dalam perangkat lunak untuk meng-identifikasi serial number atau pemesan dari program tersebut.

(3) *Consumer*

Pada sisi pengguna, ada dua aspek penting yang ingin dilindungi. Pertama, pembeli ingin meyakinkan bahwa perangkat lunak yang digunakannya tidak mengalami perubahan oleh pihak-pihak yang tidak bertanggung jawab. Kedua, pembuat perangkat lunak ingin meyakinkan bahwa pemakai tidak melanggar lisensi perjanjian, misalnya pemakai hanya menggunakan sesuai dengan lisensi yang disepakati sebelumnya.

**Validity Mark (VM)** adalah watermark yang menempelkan informasi dalam perangkat lunak untuk meyakinkan bahwa perangkat lunak tersebut tetap sama dengan aslinya ketika dibuat.

**Licensing Mark** adalah watermark menempelkan informasi dalam perangkat lunak yang mengontrol bagaimana perangkat lunak tersebut dapat digunakan.

## 2.3.2 Serangan Atas Software Watermarking [COL-2000]

*Resilient* software watermarking (seberapa tahan suatu kode watermark dari pengrusakan atau penghilangan) sangat ditentukan oleh jenis serangan yang mungkin terjadi. Dengan kata lain, berapa tingkat serangan (yang wajar) yang mungkin terjadi atau teknik apa yang mungkin digunakan oleh *attacker*.

Harus disadari bahwa secara umum dapat dikatakan bahwa tidak ada teknik software watermarking yang mampu bertahan selamanya dari *manual attack* (program diinspeksi oleh manusia secara manual).

Secara garis besar, serangan atas teknik software watermarking terbagi :

- *Additive attack.*

  Pelaku pembajakan menambahkan kode watermark lain dalam aplikasi yang telah ter-watermark sehingga akan sulit menentukan pemilik pertama.

  Pada gambar 2.6 (a), *B* menambahkan kode $W_1$ sebagai watermarknya dalam program *P'* milik *A* yang terwatermark *W*. *A* akan mengalami kesulitan dalam mengklaim bahwa *P''* adalah miliknya sebab watermark dari *B* juga terdapat di dalamnya .

- *Distortive attack.*

  Pembajak berusaha untuk merubah aplikasi dengan harapan kode watermark menjadi rusak sehingga tidak dapat dikenali lagi.

  Pada gambar 2.6 (b), *B* memberikan distortive attack pada program *P'* yang ter-watermark *W* milik *A* sehingga menyebabkan *W* berubah menjadi *W'*.

- *Collusive attack.*

  Pembajak membandingkan dua atau lebih salinan aplikasi yang mempunyai kode fingerprint berbeda sehingga dapat menemukan lokasi fingerprint dan membuangnya.

  Pada gambar 2.6 (c), *B* membeli beberapa salinan aplikasi *P* milik *A* dan membandingkan kode fingerprint didalamnya sehingga lokasi fingerprint dapat dihilangkan.

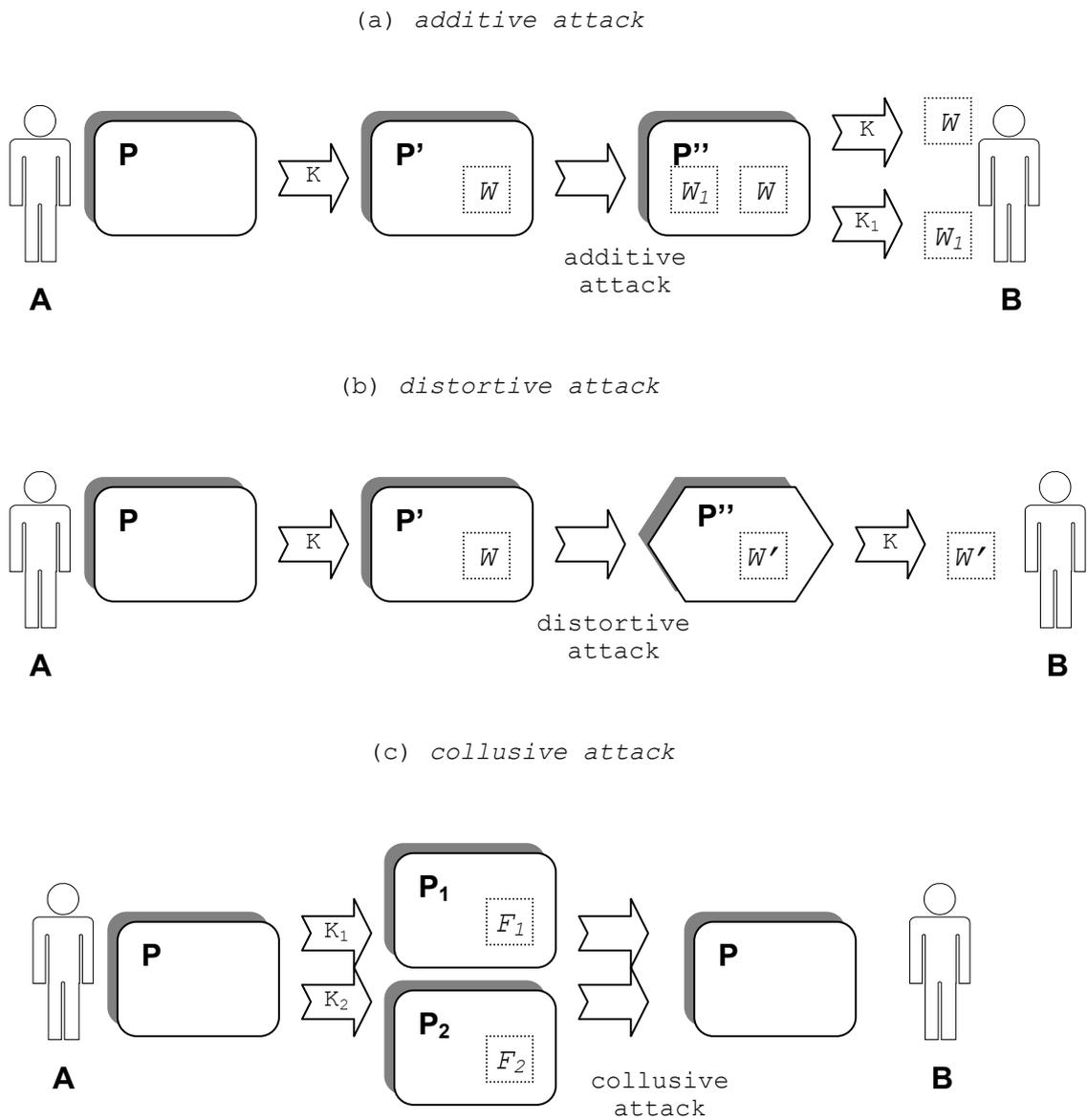

*Gambar 2.6 Jenis serangan pada software watermarking*

Diasumsikan bahwa serangan utama adalah distortive attack dalam bentuk transformasi kode. Teknik software watermarking yang diteliti diharapkan sukses melewati proses *translation* (kompilasi, dekompilasi), *optimization*, dan *obfuscation*.

## 2.4 Dummy Method Watermarking [MON-2000]

Teknik software watermarking terbagi atas dua jenis yaitu *static* dan *dynamic* watermarking. Pada teknik static watermarking, watermark diletakkan langsung dalam program, sedangkan pada dynamic watermarking kode akan terbentuk pada saat run-time dan diletakkan dalam dynamic-state program.

Dummy method watermarking termasuk static watermarking dimana kode watermark langsung ditambahkan dalam program (dalam hal ini class-file) pada suatu method "dummy". Dengan menempelkan kode yang menjadi tanda kepemilikan dalam setiap class file akan meyakinkan hak kepemilikan dari setiap class file tersebut.

Hasil penelitian Akito Monden [MON-2000] menunjukkan karakteristik dari teknik ini yaitu :

1. Watermark dalam program Java tidak mengurangi efesiensi eksekusi
2. Watermark dalam program Java dapat diperoleh hanya jika pembuat program asli menggunakan watermark-dekoding tool
3. Pengguna program sulit mengetahui lokasi watermark, sehingga menghapus dan atau merubah watermark menjadi sangat sulit
4. Termasuk jika hanya sebagian program yang di-curi dan dimasukkan dalam program lain, watermark tetap dapat di-dekoding dimanapun dia berada dalam program baru tersebut.
5. Sebagian besar watermark yang ditempelkan dalam class file tetap sukses melawan dua jenis serangan yang akan menghapus watermark: *obfuscactor attack*, dan *decompile-recompile attack*.

## 2.4.1 Watermarking Encoding

Watermarking enkoding merupakan proses memasukkan kode watermark dalam program yang ingin dilindungi. Proses ini dimulai dari proses persiapan program dan kode watermark sampai dihasilkannya program ter-watermark.

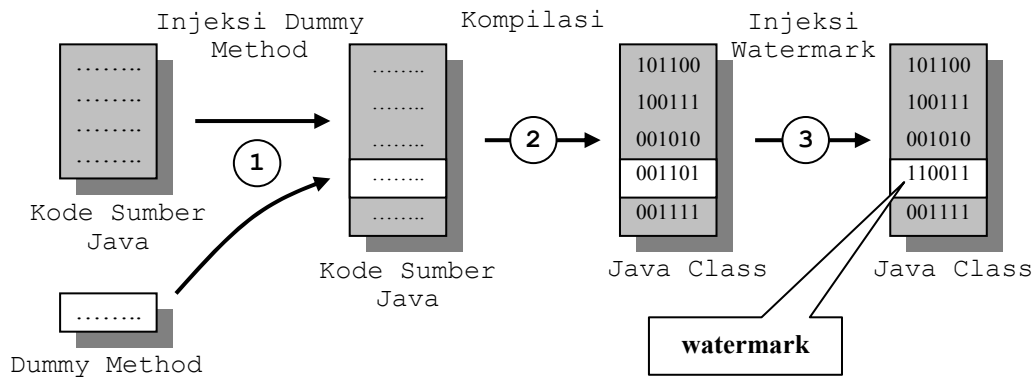

*Gambar 2.7 Tahapan-tahapan enkoding watermark
pada teknik Dummy Method Watermarking*

Tahapan-tahapan enkoding ini dapat dijelaskan sebagai berikut :

(1) Injeksi Dummy Method

Pada tahap ini satu atau beberapa method kosong (*dummy*) ditambahkan dalam kode sumber program. *Dummy method* ini berguna sebagai tempat kode *watermark* nantinya, yang tidak akan pernah dieksekusi oleh program.

Untuk menghubungkan *dummy method* dengan method-method lain dalam program ini dapat digunakan perintah :

*if (kondisi) dummyMethod();*

dimana kondisi tidak akan pernah bernilai *true*.

Dummy method harus tidak pernah dieksekusi karena pada tahapan injeksi watermark, bytecode dalam method tersebut akan diubah.

(2) Kompilasi

Tahapan selanjutnya adalah melakukan kompilasi pada kode sumber yang telah berisi *dummy method* tersebut. Proses kompilasi sama dengan proses kompilasi program yang tidak akan di-watermarking.

(3) Injeksi Watermark

Pada tahap ini kode *watermark* dimasukkan pada *class file* (dapat secara manual maupun otomatis).

Kode ini tidak dapat dimasukkan begitu saja, karena *JVM (Java Virtual Machine)* akan melakukan pengecekan *sintax* maupun konsistensi tipe data [LIN-1999].

Untuk menjaga supaya *sintax* program tetap benar dan tipe data tetap konsisten, kode *watermark* dimasukkan dalam program dengan :

a. *Overwriting Numerical Operands*

Suatu operator numerik pada sebuah *opcode* yang meng-*push* nilai pada *stack* dan pada *opcode* yang menambah nilai pada *stack* dapat di-*overwrite*. Misalnya operator 'xx' pada *opcode* 'iinc xx' dan 'bipush xx' dapat di-*overwrite* dengan data 1 byte.

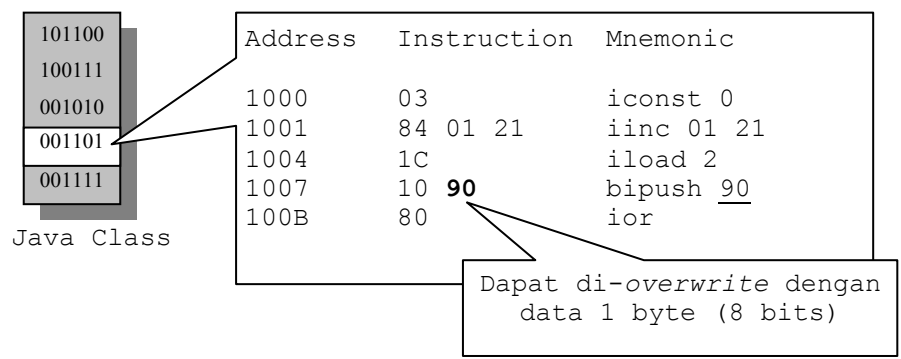

Gambar 2.8 *Overwriting numerical operands*

b. *Replacing opcodes*

Beberapa *opcode* seperti *iadd, ifnull, iflt* dapat dirubah ke opcode lain. Misalnya mengganti *iadd* menjadi *isub* tidak akan menyebabkan kesalahan *sintax* ataupun melanggar konsistensi tipe data. Opcode *iadd* dapat diganti dengan *isub, imul, idiv, irem, iand, ior* dan *ixor* sehingga kita dapat mengkodekan 3 bit informasi dalam opcode-opcode ini. Misalnya, 000 ke *iadd*, 001 ke *isub*, 010 ke *imul*, ..., 111 ke *ixor*.

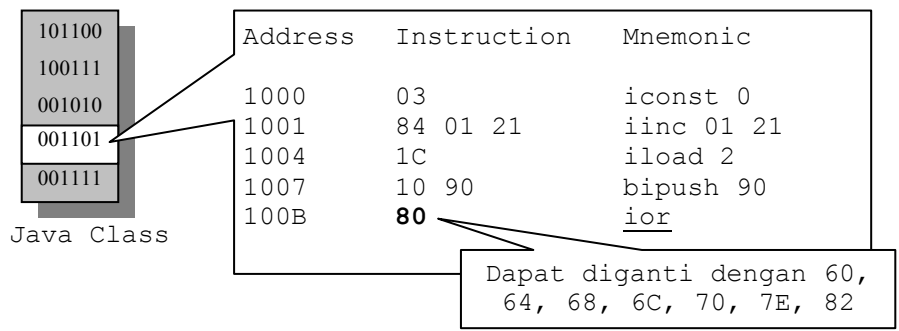

*Gambar 2.9  Replacing opcodes.*

Kedua proses ini tidak akan merubah karakteristik dari *class file*, karena dilakukan pada *dummy method* yang tidak akan pernah dikerjakan.

Sebagai contoh, untuk meng-*enkode* 'ITS SURABAYA' :

- Kode watermark dirubah menjadi urutan bit sesuai dengan aturan pengkodean bit (gambar 2. 10 (a)).
- Membuat aturan pengkodean bit dalam opcode atau numerik (gambar 2.10 (b)).
- Bit-bit dari kode watermark selanjutnya dapat di-kode-kan pada program (gambar 2.10 (c)).

(a) Contoh aturan peng-kodean dalam urutan bit dan hasil
    peng-kodean 'ITS SURABAYA'

```
Character        Bit Code

Space    ...     0000
I        ...     0001
T        ...     0010
S        ...     0011
U        ...     0100
R        ...     0101
A        ...     0110
B        ...     0111
Y        ...     1111
```

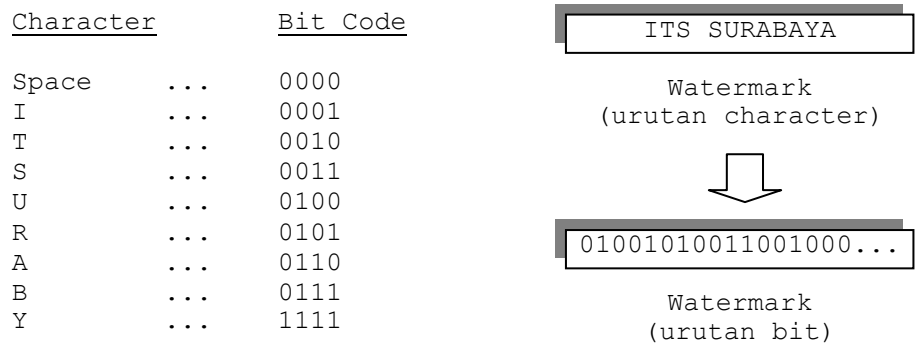

(b) Contoh aturan peng-kodean bit dalam instruksi

```
    Instruction         Assigned bits

60  iadd    ...         000
64  isub    ...         001
68  imul    ...         010
6C  idiv    ...         011
70  irem    ...         100
7E  iand    ...         101
80  ior     ...         110
82  ixor    ...         111

9B  iflt    ...         00
9D  ifge    ...         01
9C  ifgt    ...         10
9E  ifle    ...         11

C6  inull    ...        0
C7  inonnull ...        1
```

(c) Contoh proses enkoding

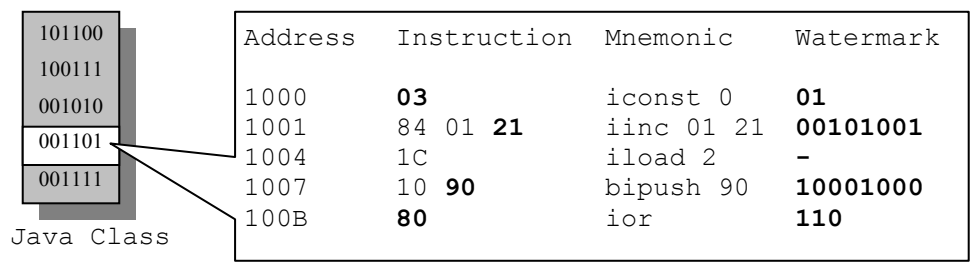

*Gambar 2.10 Contoh enkoding kode 'ITS SURABAYA'*

**2.4.2 Watermarking Decoding**

Proses decoding watermark dimaksudkan untuk mendapatkan kembali kode yang sebelumnya telah di-encodekan didalamnya. Kalau pada proses encoding, nama atau letak dummy method diketahui dengan pasti, pada proses ini hal tersebut tidak berlaku lagi. Hal ini disebabkan karena ada kemungkinan terjadi perubahan pada nama atau letak dummy method (misalnya, karena sudah dikenai attack).

Proses decoding memperhatikan :

- Hubungan antara bytecodes dengan bit-bit kode (gambar 2.10 (b)).
- Hubungan antara urutan bit dengan karakter (gambar 2.10 (a)).

Dilakukan dengan cara :

- Mengganti operand-operand dan opcode-opcode kedalam urutan bit sesuai dengan aturan pengkodean (gambar 2.11 (a)).
- Mengganti urutan bit tersebut dalam urutan character (gambar 2.11 (b)).

(a) Mengganti operand dan opcode menjadi urutan bit sesuai dengan aturan pengkodean.

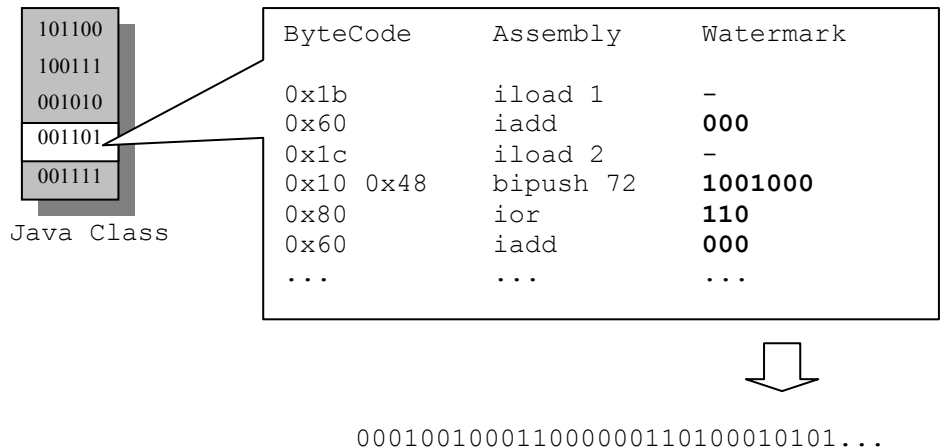

(b) Mengganti urutan bit menjadi karakter sesuai
    dengan aturan pengkodean karakter.

```
0001  =  I
0010  =  T
0011  =  S
0000  =  space
0011  =  S
0100  =  U
0101  =  R
01..  =  ...
```

*Gambar 2.11 Contoh proses watermark dekoding*

## 2.5  Opaque Predicates [COL-1999]

Opaque predicates adalah salah satu teknik obfuscation untuk control-flow transformation. Obfuscation merupakan proses untuk men-transformasi program *A* menjadi *A'* sedemikian sehingga :

- Program *A'* mempunyai karakteristik yang ekuivalen dengan *A*.
- Program *A'* lebih sulit untuk dimengerti.

Konsep dasar obfuscation transformation adalah :

Misalkan terdapat sejumlah obfuscating transformations $T = \{T_i, ..., T_j\}$, dan program *P* yang terdiri atas obyek-obyek kode sumber (class, method, statement, dll.), $\{S_1, ..., S_k\}$, dibuat suatu program baru $P' = \{..., S'_j = T_i(S_j), ...\}$ sedemikian sehingga:

- *P'* mempunyai sifat (*behavior*) yang sama dengan *P*, yaitu transformasi bersifat *semantics-preserving*,

- *obscurity P'* di-maksimal-kan, yaitu untuk mengerti dan me-reverse engineering *P'* membutuhkan waktu lebih banyak dibandingkan untuk mengerti dan me-reverse engineering *P*,

- *resilience* untuk setiap transformasi $T_i(S_j)$ di-maksimal-kan, yaitu sulit untuk membuat suatu tool otomatis untuk mengembalikan hasil transformasi, ataupun eksekusi tool ini membutuhkan waktu yang sangat lama,
- *stealth* dari setiap transformasi $T_i(S_j)$ di-maksimal-kan, yaitu *statistical properties* dari $S'_j$ adalah serupa untuk semua $S_j$,
- *cost* (waktu eksekusi /space penalty yang digunakan proses transformasi) dari *P'* di-minimal-kan.

Untuk pengertian opaque predicates sendiri dengan mengacu pada peneliti dalam bidang software watermarking antara laian Collberg dan Thomborson menyatakan "*A predicate P is opaque if its outcome is known at obfuscation time, but is difficult for deobfuscator to deduce.*" Sementara Douglas Low [LOW-1998A] menyatakan "*A boolean valued expression whose value is known to an obfuscator, but is difficult for deobfuscator to deduce.*"

Control-flow transformation memecah suatu aliran prosedur program dengan menyisipkan suatu predicate *P*, dimana *P* akan selalu bernilai *true* (ditulis $P^T$) atau *P* selalu bernilai *false* (ditulis $P^F$) atau dapat bernilai *true* ataupun *false* (ditulis $P^?$).

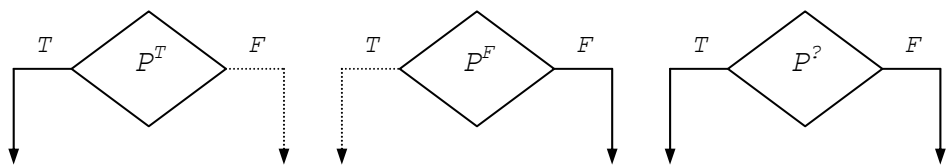

*Gambar 2.12 Jenis-jenis opaque predicates. Garis putus-putus mengindikasikan jalur yang tidak akan pernah dikerjakan.*

Transformasi ini menyisipkan *opaque predicates* yang akan membuat program menjadi sulit dievaluasi dengan *deobfuscator*. Contoh penerapan obfuscation dengan opaque predicates dapat dilihat pada gambar 2.13.

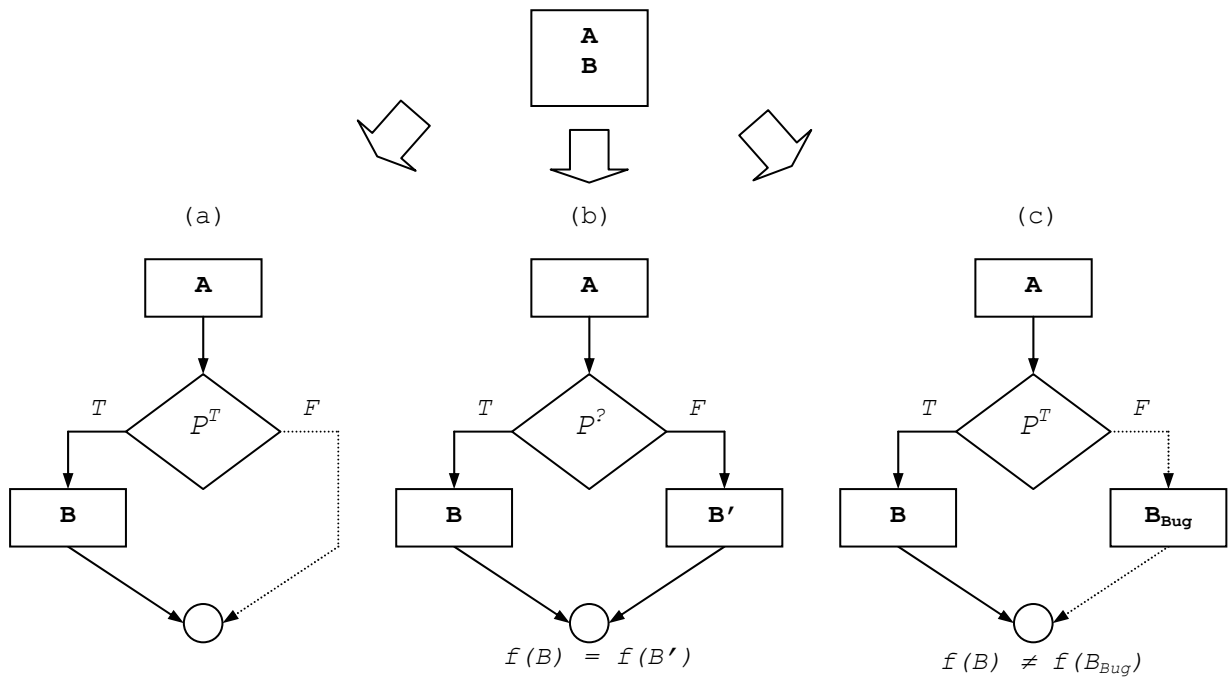

*Gambar 2.13 Contoh penerapan transformasi control-flow dengan opaque predicates*

Pada gambar 2.13 (a) blok dipecah dengan menyisipkan *opaquely true* predicate $P^T$ yang akan selalu dievaluasi True. Pada gambar (b), blok dipecah dalam versi *B* dan *B'* yang merupakan dua versi hasil *obfuscation* yang berbeda, yang dalam kenyataannya melakukan kerja yang sama. *Predicate* $P^?$ memilih salah satu dari mereka pada *runtime*. Sementara gambar (c), blok dipecah menjadi *B* dan $B_{Bug}$ (versi *bug* dari *B*). $P^T$ akan selalu memilih versi yang benar *B*.

Untuk membuat opaque predicates dapat dilakukan dengan berbagai cara baik yang bersifat static maupun dinamis. Untuk opaque predicate yang statis, bentuk opaque predicate tersebut selalu tetap, walaupun variabel-variabel yang ada didalamnya dapat berubah-ubah dan nilai yang dihasilkan opaque predicate ini juga selalu tetap.

Contoh opaque predicate statis :

$$(7X^2 - 1) == Y^2$$

Untuk X dan Y adalah sembarang bilangan bulat (0 … ∞), formula ini akan selalu menghasilkan *false*.

Dalam penelitian ini, karena yang akan digunakan adalah opaque predicate yang bersifat dinamis, maka pembentukan opaque predicate-nya juga yang bersifat dinamis, yaitu :

**Object and aliases [COL-2002] :**

Defenisi: Aliasing akan terjadi jika dua variabel menunjuk pada lokasi memori yang sama. Aliasing terdapat pada bahasa yang mempunyai referensi parameter, pointer atau array.

Dapat dilakukan dengan:

1. Membuat dua atau lebih stuktur data (*object*) $S_1, S_2, ...$

2. Membuat sejumlah pointer $P_1, P_2, ...$ yang menunjuk pada $S_1, S_2, ...$

3. Membuat program untuk meng-update stuktur data ini dengan aturan-aturan tertentu, misalnya "$P_1$ tidak akan pernah menunjuk pada lokasi memori yang ditunjuk oleh $P_3$".

4. Struktur data ini (dengan variasi-nya) dapat digunakan untuk membentuk opaque predicates.

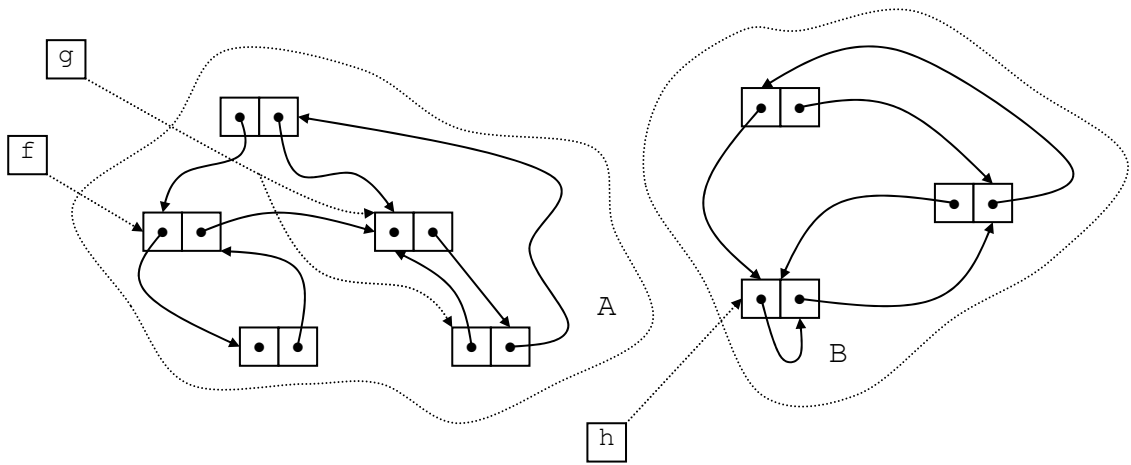

*Gambar 2.14 Contoh object and aliases.*
*Untuk (f == g) akan membentuk opaque predicates $P^?$*

Pada gambar 2.14, terdapat dua buah struktur data (*A* dan *B*). `f` dan `g` adalah pointer yang menunjuk pada *A*, sedangkan `h` menunjuk *B*. Jika diasumsikan bahwa pointer dalam *A* tidak bisa berpindah ke *B* dan sebaliknya, hanya berpindah dalam struktur data masing-masing, maka dapat dibentuk opaque predicate (`f == g`)$^?$, yaitu nilainya bisa *true* atau *false*. Akan bernilai *false* jika menunjuk *node* yang berbeda dan *true* jika menunjuk *node* yang sama (`f` dan `g` berada dalam struktur data yang sama). Untuk (`g == h`) akan membentuk $P^F$ sebab `g` dan `h` tidak akan pernah mungkin menunjuk *node* yang sama sebab berada dalam struktur data yang berbeda.

**Concurrency [COL-2002] :**

Dapat dilakukan dengan:

1. Membuat suatu struktur data global *V*.

2. Membuat beberapa *thread* yang dalam selang waktu tertentu akan meng-update stuktur data *V* secara *concurrency*.

3. Opaque predicates dapat dibentuk dari data dalam *V*.

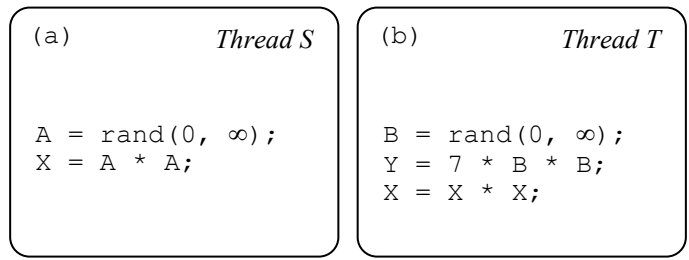

*Gambar 2.15 Contoh concurrency.*
*Untuk ((Y - 1) == X) akan membentuk opaque predicates $P^F$*

Gambar 2.15 memperlihatkan dua buah thread yaitu *S* (a) dan *T* (b). Kedua thread ini bersama-sama membentuk opaque predicate $(7X^2 - 1) == Y^2$ yang bernilai *false*. *S* meng-update nilai X sedangkan *T* meng-update nilai Y, dimana nilai X, Y diperoleh dari bilangan yang dibangkitkan secara acak dari 0 sampai ∞.

Contoh ini menggunakan statis opaque predicate $(7X^2 - 1) == Y^2$ yang dimodifikasi sehingga nilai X dan Y bersifat dinamis. Walaupun terlihat dinamis, rumus yang digunakan tetap statis.

**Concurrency and Aliasing [COL-2002] :**

Merupakan penggabungan teknik *object and aliasing* dengan *concurrency*, dilakukan dengan:

1. Struktur data *V* merupakan stuktur data dinamis.
2. Setiap thread dihubungkan masing-masing satu pointer pada *V*.
3. Secara *synchronous*, *thread-thread* memindahkan pointer-nya masing-masing.

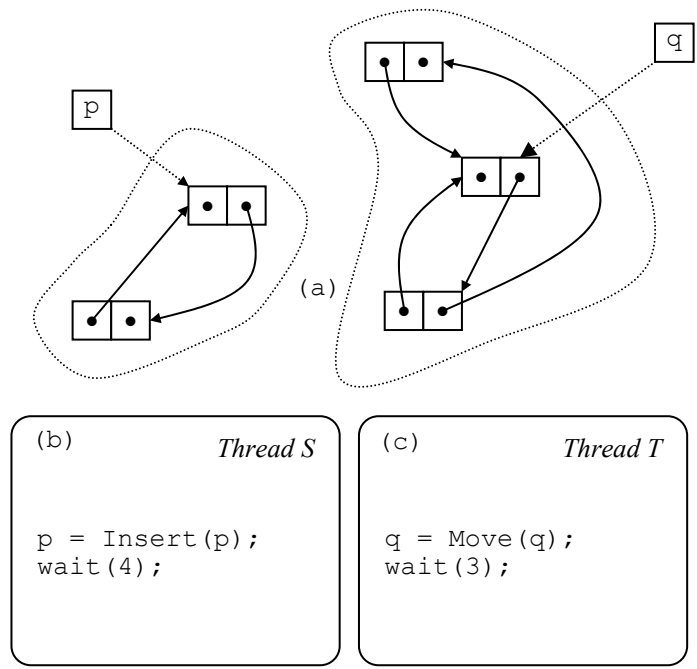

*Gambar 2.16 Contoh concurrency dan Aliasing.
Untuk (p == q) akan membentuk opaque predicates $P^F$*

Pada gambar 2.16, terdapat dua struktur data (a) dengan pointer `p` dan `q`. Juga dibuat thread *S* (b) dan *T* (c), dimana *S* meng-update pointer `p` sedangkan *T* meng-update pointer `q`. Untuk (`p == q`) akan mempunyai nilai *false* karena `p` tidak akan pernah menunjuk alamat sama dengan `q` (berada pada struktur data yang berbeda).

## 2.6 Dynamically Opaque Predicates [PAL-2000]

Secara garis besar, tindakan *attacker* untuk mengganggu, merusak atau menghilangkan kode watermark dalam program dapat dikategorikan menjadi :

(i) *Static analisis*, yaitu meng-analisis program pada saat program tidak dijalankan. Hal ini dapat dilakukan jika *attacker* dapat memperoleh kode sumber misalnya dengan menggunakan *decompiler* untuk meng-*generate* kode sumber tersebut.

(ii) *Dynamic analisis*, yaitu menganalisa program pada saat dijalankan. Hal ini dapat dilakukan dengan menggunakan tool-tool untuk memonitor heap ataupun memori ketika program dijalankan dengan input yang bervariasi sehingga isi maupun perubahan yang terjadi pada heap atau memori dapat diketahui.

Untuk kondisi yang terburuk, seorang *attacker* dapat mempunyai akses pada kode sumber program maupun heap dan memori pada saat program yang telah di-watermark dijalankan. *Attacker* semacam ini dikenal dengan "*expert attacker*" [PAL-2000].

Salah satu permasalahan yang bisa timbul pada opaque predicates yaitu jika seorang attacker dapat memonitor heap, register, dll. selama eksekusi program, maka terdapat kemungkinan untuk mengetahui suatu predicate selalu bernilai *true* atau *false* [PAL-2000]. Untuk menjaga hal ini predicate dapat dibuat untuk mempunyai nilai yang berlainan pada waktu eksekusi yang berbeda.

Penggunaan opaque predicate yang dinamis mempunyai tujuan utama :
1. Membuat kode watermark tidak dapat dikenali.
2. Membuat kode program asli dan watermark tidak dapat dibedakan.

Jika kedua hal ini dapat diterapkan maka seorang *expert attacker*-pun akan kesulitan untuk mengganggu program dengan watermark-nya [PAL-2000].

Ide dasar:
1. Mengelompokkan beberapa opaque predicates tertentu.
2. Setiap opaque predicate dalam setiap kelompok dapat mempunyai nilai yang berbeda pada waktu eksekusi yang lain.
3. Opaque predicates yang dihasilkan berdasarkan nilai kelompok secara keseluruhan.

| Predicates | Run 1 | Run 2 | Run 3 | Run 4 |
|---|---|---|---|---|
| $P_1$ | T | F | F | F |
| $P_2$ | F | T | T | F |
| $P_3$ | T | T | F | F |
| $P_4$ | F | F | T | T |

*Gambar 2.17 Contoh pengelompokan opaque predicates. Opaque predicates yang dihasilkan berdasarkan nilai kelompok secara keseluruhan*

# BAB III

# DESAIN DAN IMPLEMENTASI SYSTEM

## 3.1 Desain Teknik Software Watermarking

Beberapa hal utama yang perlu diperhatikan dalam mendesain suatu teknik software watermarking [COL-1999, COL-2000] :

(i) *Required Data Rate*, berapa besar kode watermark dibandingkan dengan ukuran program yang ingin dilindungi.

Pada dummy method watermarking, kode watermark diletakkan pada suatu method kosong sehingga besar kode watermark adalah berdasarkan besar method tersebut.

(ii) *Form of Cover Program*, bentuk program yang ingin dilindungi, apakah program akan didistribusikan dalam *architecture-neutral virtual machine code type* atau *untyped native binary code.*

Penelitian ini dilakukan pada program Java yang bersifat *architecture-neutral virtual machine code*, namun demikian bukan tidak mungkin teknik yang digunakan ini juga dapat diterapkan pada *untyped native binary code.*

(iii) *Expected Thread-Model*, jenis watermarking attack yang mungkin terjadi pada program.

Sebagaimana telah dijelaskan pada Bab II sebelumnya, dari tiga jenis watermarking attack yang mungkin terjadi, diasumsikan bahwa *distortive attack* merupakan attack utama yang dapat berupa transformasi kode. Sistem diharapkan sukses melewati proses *translation* (kompilasi, dekompilasi, *optimization*, dan *obfuscation*.

## 3.2 Desain Dummy Method

Method "dummy" adalah sama dengan method-method lain dalam class-file kecuali pada fungsi yang dikerjakan. Method umumnya berisi data atau proses tertentu yang ada hubungannya dengan aplikasi, sedangkan method dummy ini tidak mempunyai hubungan dengan aplikasi sama sekali.

Resistansi teknik watermarking ini sangat ditentukan oleh metode injeksi watermark yaitu *Overwriting Numerical Operands* dan *Replacing Opcodes*. Kedua metode ini dapat diterapkan secara tunggal ataupun bersama-sama. Hasil penelitian Akito Monden [MON-2000] menunjukkan bahwa *Overwriting Numerical Operands* lebih tahan terhadap attack dibandingkan dengan *Replacing Opcodes* ataupun kombinasinya.

### 3.2.1 Bentuk dan Jumlah Dummy Method

Karena dummy method merupakan tempat kode, maka besar-nya sangat tergantung pada besar kode watermark. Semakin besar kode maka semakin besar pula dummy method yang digunakan, yang juga berarti akan membuat kode sumber semakin besar. Akito Monden [MON-2000] menunjukkan bahwa untuk membuat tempat kode yang besar dengan ukuran kode sumber yang kecil, dummy method sebaiknya berisi operasi perhitungan matematik.

Demikian pula dengan jumlah dummy method dalam program tergantung berapa kali kode watermark akan diikutkan dalam program. Semakin sering kode diikutkan (dengan harapan program semakin aman) berarti semakin besar kode sumber.

```
private void X(int k){
   int i, j;
   for(i = 0; i < 10 ; i++)
       for(j = 0; j < 10 ; j++) k+=i*10+j;
   System.out.println("k = " + k);
   for(i = 0; i < 20 ; i++)
       for(j = 0; j < 30 ; j++) k+=i*3-j;
   System.out.println("k = " + k);
   for(i = 0; i < 25 ; i++)
       for(j = 0; j < 20 ; j++) k+=i*4-j*3;
   System.out.println("k = " + k);
}
```

*Gambar 3.1 Contoh Dummy Method*

### 3.2.2 Penggunaan Kunci

Pada umumnya kode watermark merupakan urutan huruf atau angka yang mempunyai arti bagi pembuatnya. Untuk merubah bentuk watermark, dapat digunakan kunci dengan harapan akan lebih mempersulit mengenali kode tersebut. Hal ini dilakukan dengan meng-kodekan kunci kemudian menambahkannya pada kode watermark dengan operasi tertentu.

```
Kode      :   110011001000110
Kunci     :         1100101011
                                  AND
Kode Baru :   110011000000010
```

*Gambar 3.2 Contoh Penggunaan Kunci dengan operator AND*

### 3.3 Desain Dynamically Opaque Predicates

Karena opaque predicates merupakan nilai *boolean*, maka kondisi-kondisi *boolean* pada program asli juga dapat diikutkan pada pengelompokan sehingga akan menyulitkan *attacker* untuk membedakan apakah opaque predicate yang ada tersebut adalah salah satu dari :

- program asli,

- atau traditional opaque predicate (mempunyai nilai tetap),

- atau dynamically opaque predicate.

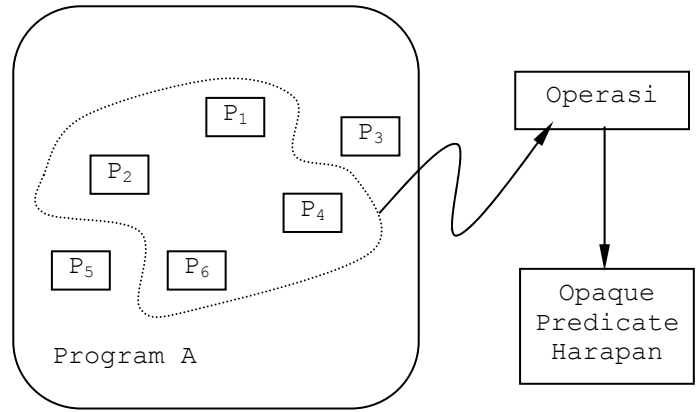

*Gambar 3.3 Proses pengelompokan opaque predicates untuk menghasilkan opaque predicate yang diinginkan*

Pada gambar 3.3, dalam `Program A` terdapat 6 opaque predicate ($P_1, ..., P_6$). Dibuat kelompok yang terdiri dari $P_1, P_2, P_4, P_6$ yang diberikan `operasi` sehingga diperoleh opaque predicate harapan.

Dalam hubungannya dengan teknik *dummy method*, opaque predicate yang akan dihasilkan adalah harus selalu bernilai *false* (dummy method berisi kode watermark, dimana bytecode method ini sudah dirubah). Untuk itu harus dibuat sedemikian rupa sehingga nilai-nilai dari beberapa opaque predicate tersebut secara keseluruhan akan tetap bernilai *false*.

Kunci utama dalam dynamically opaque predicates yaitu pada pengelompokan opaque predicate dengan memperhatikan operasi yang akan digunakan, yang mana harus selalu menghasilkan nilai kelompok *false*.

| Predicates | Run 1 | Run 2 | Run 3 | Run 4 |
|---|---|---|---|---|
| $P_1$ | T | F | F | T |
| $P_2$ | F | T | T | T |
| $P_3$ | T | T | F | T |
| **AND** | **F** | **F** | **F** | **T** |

*Gambar 3.4  Contoh pengunaan operator AND. Ada kemungkinan menghasilkan nilai yang tidak diharapkan*

### 3.3.1  Desain Pengelompokan Opaque Predicates

Untuk menjaga agar supaya hasil tetap sesuai dengan yang diinginkan, ada beberapa kemungkinan yang dapat dilakukan :

(i) Anggota kelompok tetap. Untuk operasi AND, supaya nilai kelompok selalu *false* berarti harus ada anggota minimal satu yang bernilai *false* dalam setiap *run*-nya.

(ii) Anggota kelompok dinamis. Pengelompokan dilakukan dengan selalu memasukkan minimal satu opaque predicate yang bernilai *false* dalam kelompok dalam setiap *run*-nya.

Dengan memperhatikan karakteristik software watermarking (*resilient, high data rate, cheap, stealthy*) sebagaimana dijelaskan pada Bab II, penggunaan cara pertama lebih mudah diterapkan, namun tidak berarti bahwa cara kedua tidak dapat diterapkan.

Untuk pengelompokan dengan anggota kelompok dinamis dapat diharapkan tingkat resilient yang lebih baik, namun lebih rumit dalam penerapannya. Penelitian ini dibatasi pada anggota kelompok tetap.

**Algoritma I : Kelompok Tetap, Tidak Bersyarat**

1. Memilih opaque predicate $P_1, ..., P_n$ untuk dijadikan kelompok, yang terdiri dari opaque predicate yang dinamis termasuk kondisi-kondisi *boolean* program yang memungkinkan.

2. Terdapat satu opaque predicate traditional $P_1$ yang bernilai *false* ($P^F$), $P_2 ... P_n$ dapat bernilai apa saja ($P^?$).

3. Sehingga dapat diperoleh : $\sum_{i=0}^{n} P_i \approx false$

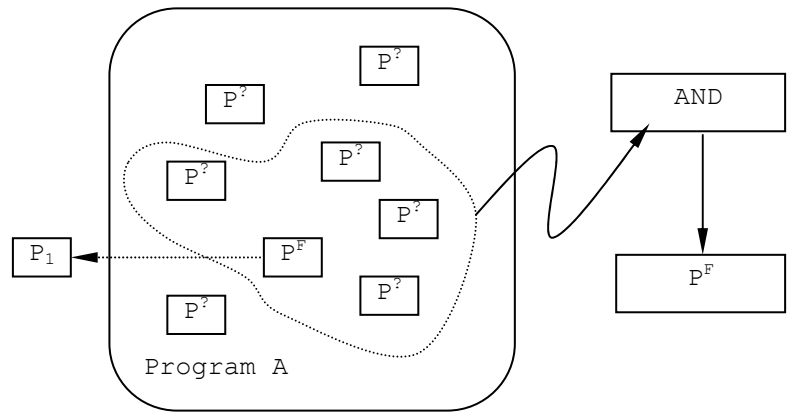

*Gambar 3.5  Contoh pengelompokan Kelompok Tetap Tidak Bersyarat dengan operator AND*

Pada gambar 3.5, dalam `Program A` terdapat beberapa opaque predicate $P^?$ dan satu $P^F$. Untuk pengelompokan algoritma I ini, $P^F$ dipilih menjadi $P_1$ sebagai acuannya, yang diikutkan dalam kelompok. Dengan demikian dengan operator `AND` nilai kelompok ini pasti *false*. Dengan $P_1$ bernilai *false*, maka opaque predicate lainnya dapat bernilai apa saja ($P^?$).

**Algoritma II : Kelompok Tetap, Bersyarat**

1. Memilih opaque predicate $P_1, ..., P_n$ untuk dijadikan kelompok, yang terdiri dari opaque predicate yang dinamis termasuk kondisi-kondisi *boolean* program yang memungkinkan.

2. Terdapat dua opaque predicates yang dinamis sebagai acuan $P_1$ dan $P_2$.

3. Jika $P_1$ bernilai *true* ($P_1^T$), maka $P_2$ bernilai *false* ($P_2^F$), dan jika $P_1$ bernilai *false* ($P_1^F$), maka $P_2$ dapat bernilai apa saja ($P_2^?$). $P_3 ... P_n$ dapat bernilai apa saja ($P^?$).

4. Sehingga dapat diperoleh : $\sum_{i=0}^{n} P_i \approx$ *false*

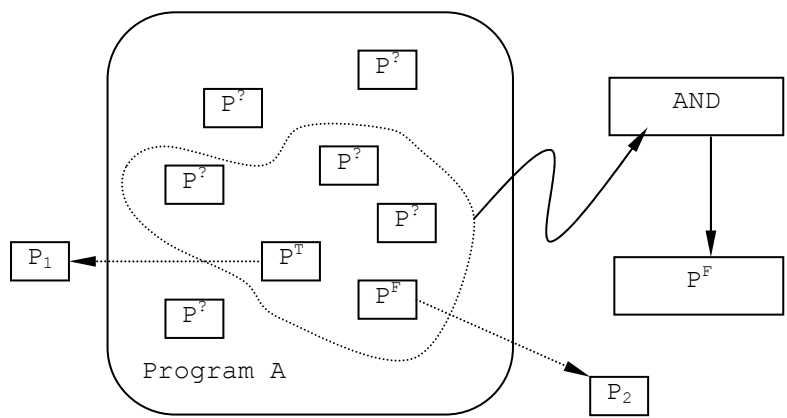

*Gambar 3.6 Contoh pengelompokan Kelompok Tetap Bersyaratat dengan operator AND*

Pada gambar 3.6, dalam `Program A` terdapat beberapa opaque predicate. Dibuat kelompok dan memilih dua anggota sebagai acuan ($P_1$ dan $P_2$). Jika $P_1$ bernilai *true*, maka $P_2$ harus bernilai *false*, dan jika $P_1$ bernilai *false* ($P_1^F$), maka $P_2$ dapat bernilai apa saja ($P_2^?$). $P_3 ... P_n$ dapat bernilai apa saja ($P^?$). Penggunaan kata "bersyarat" ini mengacu pada proses penge-cekan nilai $P_1$ yang dilakukan supaya nilai kelompok tetap *false* (jika menggunakan operator `AND`).

Teknik pengelompokan dengan algoritma I (tidak bersyarat) adalah lebih mudah untuk ditelusuri dibandingkan algoritma II (bersyarat), namun cara yang kedua lebih rumit dalam penerapannya dibandingkan cara yang pertama.

### 3.3.2 Penerapan Dynamically Opaque Predicates

Untuk menghasilkan opaque predicate yang dinamis, pembentukan-nya dilakukan dengan menggunakan teknik yang dinamis pula yaitu *Object and Aliases, Concurrency,* maupun *Concurrency and Aliasing* [COL-2002].

Pembentukan opaque predicate yang dinamis dilakukan secara *offline* dengan menambahkan dalam program asli kode untuk :

1. Membuat suatu struktur data *V* yang merupakan stuktur data dinamis. Kode dapat diletakkan pada awal program sehingga ketika program dijalankan, struktur data telah terbentuk.

    ```
    ...
    MyClass(int wt) {
        this.wt = wt;
        g = new Node();
        g.token = true;
        h = new Node();
        h.token = true;
        p = g.addNode();
        q = h.addNode();
    }
    ...
    ```

2. Membuat beberapa *thread* yang mengatur satu atau beberapa pointer pada *V*.

    ```
    ...
    Thread T, S;
    ...
    T = new Thread(this);
    T.start();
    S = new Thread(this);
    S.start();
    ...
    ```

3. Membuat *thread-thread* memindahkan pointer-nya masing-masing secara *synchronous*.

   ```
   ...
   Thread ct = Thread.currentThread();
   if (ct == T) {
       p = p.MoveNext();
   } else if (ct == S) {
       q = q.MoveBack();
   }
   try {
       if (ct == T) {
           Thread.sleep(2000);
       } else if (ct == S) {
           Thread.sleep(wt);
       }
   } catch (InterruptedException ie) {}
   ...
   ```

4. Membuat satu atau beberapa opaque predicate tradisional yang bernilai *false*, jika pengelompokan secara Tetap Tidak Bersyarat.

   ```
   ...
   b2 = (h.equals(p) || p.token);
   ...
   ```

5. Membuat penge-cek opaque predicate $P_1$ yang akan menge-set $P_2$ menjadi *false* apabila $P_1$ bernilai *true*, jika pengelompokan secara Tetap Bersyarat.

   ```
   ...
   if (p1) p2 = false;
   ...
   ```

6. Menjadikan nilai kelompok (sudah berbentuk dynamically opaque predicates) sebagai kondisi untuk meng-eksekusi dummy method.

   ```
   ...
   b1 = p.token;
   if (b2 && b1 && (g.equals(h))) {
       methodX(10);
   }
   ...
   ```

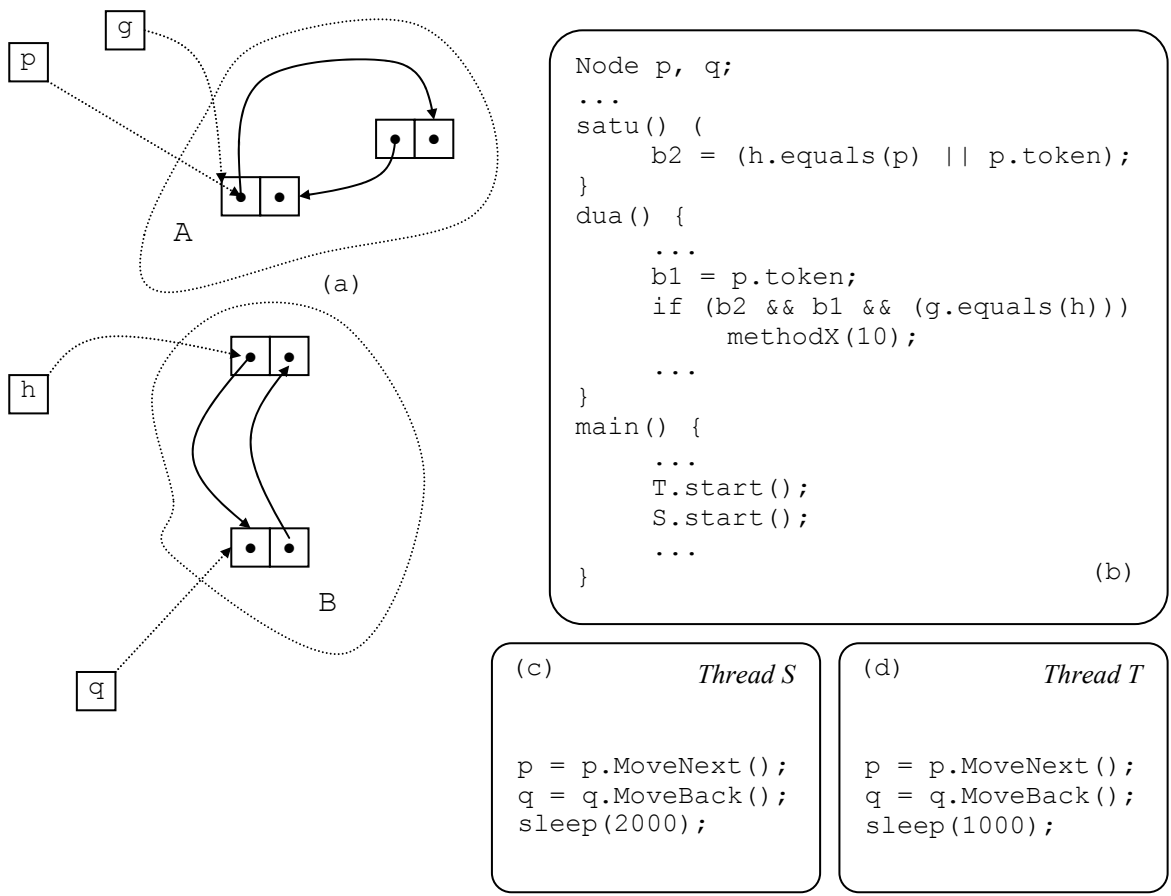

*Gambar 3.7 Contoh penerapan Dynamically Opaque Predicates menggunakan Algoritma I dengan operator AND*

Pada gambar 3.7, (a) menggambarkan dua buah struktur data *A* dan *B*. *A* mempunyai dua pointer p dan g, sedangkan pada *B* pointer h dan q. Bagian (b) merupakan kode program untuk membentuk opaque predicate termasuk dynamically opaque predicate yang akan meng-eksekusi methodX yang merupakan dummy method. *Thread* S (c) dan T (d) merupakan dua buah thread yang akan meng-update stuktur data. Thread S akan meng-update struktur data *A* setiap 2000 milidetik dan thread T untuk struktur data *B* setiap 1000 milidetik.

### 3.4 Penerapan Dummy Method dengan Dynamically Opaque Predicates

Penerapan teknik ini mempunyai dua tahapan besar, yaitu :

I. Proses Secara *Off-Line*

Proses pada tahapan ini langsung berhubungan dengan kode sumber.

(i) Memasukkan dummy method

Dummy method yang sudah dipersiapkan diikutkan pada program. Letak dalam program dapat dimana saja, jumlahnya juga tergantung kebutuhan.

(ii) Memasukkan dynamically opaque predicates

(iii) Membuat beberapa opaque predicates sesuai kebutuhan

(iv) Mencari variabel-variabel *boolean* pada program asli.

(v) Menge-lompokkan opaque predicates yang dibuat termasuk variabel *boolean* asli program dalam satu atau lebih kelompok.

(vi) Nilai kelompok-kelompok ataupun nilai keseluruhan kelompok ini menjadi kondisi untuk mengeksekusi dummy-dummy method.

II. Proses Secara *On-Line*

Source program dari tahapan pertama lalu di-*compile* seperti biasa sehingga menghasilkan java class file yang siap dieksekusi.

Kode lalu di-*encode*-kan pada class file tersebut, dengan cara :

(i) Membaca class file untuk mencari method-method dalam program (misalnya, dengan menggunakan library BCEL).

(ii) Meng-kodekan kode watermark dalam method yang dipilih seperti yang dijelaskan sebelumnya.

Proses ini menghasilkan class file baru yang telah mempunyai kode watermark.

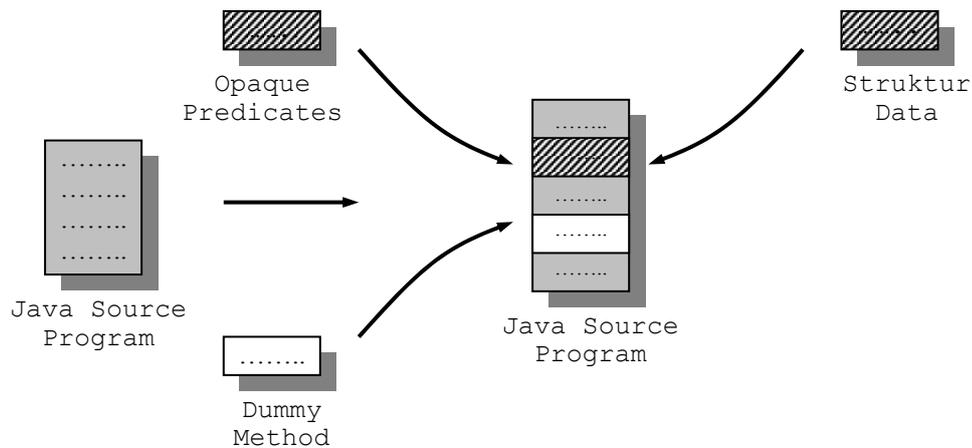

*Gambar 3.8 Tahapan memasukkan dummy method, struktur data, thread dan opaque predicates pada Java Source Program
(dilakukan secara off-line)*

Pada gambar 3.8, dummy method, struktur data, opaque predicate langsung dimasukkan dalam kode sumber. Letak, jumlah dan besar kode-kode tambahan tersebut adalah sesuai kebutuhan, sebagaimana dijelaskan sebelumnya.

Tahapan ini dilakukan pada satu class file (dummy method watermarking adalah teknik untuk melindungi Java class file). Jika aplikasi terdiri dari beberapa beberapa class file, maka tahapan ini diulang untuk setiap class file yang ingin dilindungi.

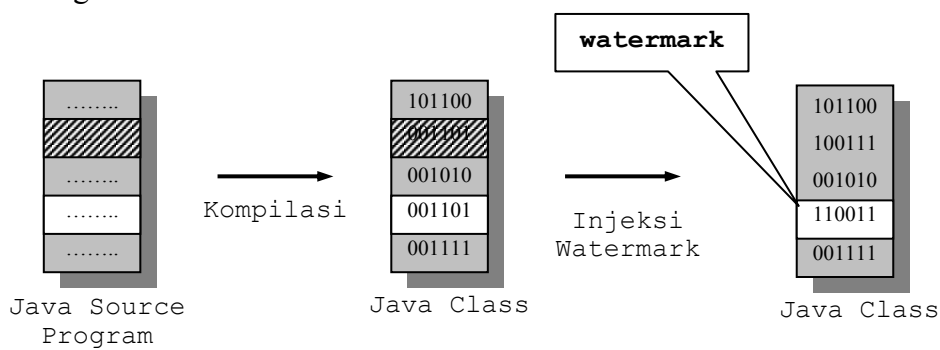

*Gambar 3.9 Tahapan kompilasi dan peng-injeksian kode watermark
(dilakukan secara on-line)*

Penekanan penelitian ini adalah pada penerapan dynamically opaque predicates (masuk pada tahapan off-line). Sementara pada proses on-line, tahapan injeksi watermark adalah menggunakan algoritma Akito Monden [MON-2000] yang diimplementasikan dalam program *jmark*, dengan beberapa modifikasi :

(i) Program jmark ditulis dalam bahasa C, pada penelitian ini dirubah menggunakan bahasa Java.

(ii) Pada program jmark, posisi dummy method langsung diikutkan sebagai parameter untuk meng-eksekusi program. Pada penelitian ini, program dimodifikasi sehingga nama method langsung dapat diketahui oleh pengguna. Modifikasi ini dilakukan dengan menggunakan library BCEL, sehingga kemungkinan terjadinya kesalahan dalam penentuan dummy method dapat diminimalisasi.

# BAB IV

# PENGUJIAN DAN ANALISA SISTEM

Untuk meng-evaluasi kualitas suatu skema watermarking, terlebih dahulu harus diketahui dengan baik jenis-jenis serangan (attack) yang mungkin terjadi. Pada umumnya, tidak ada skema yang dapat bertahan dari semua attack [COL-1999], dan sering juga beberapa teknik digunakan secara bersama-sama untuk memperoleh tingkat ketahanan yang lebih baik.

Attack yang akan diujicobakan pada sistem adalah *translation* (seperti kompilasi, dekompilasi dan translasi biner), *optimization* dan *obfuscation*. Sehubungan dengan hal tersebut, beberapa pertanyaan yang menjadi acuan dalam pengujian sistem adalah :

(i) Pada struktur data apa sebaiknya watermark diletakkan.

(ii) Bagaimana kita meng-ekstrak watermark dan meng-klaim bahwa itu adalah milik kita.

(iii) Bagaimana kita mencegah seorang attacker merusak watermark.

(iv) Seberapa besar watermark mempengaruhi performance program.

Perlu diketahui bahwa masalah yang dihadapi pada software watermarking mempunyai perberbedaan dengan yang muncul dalam media watermarking. Sebagai contoh, sulit untuk melawan *collusive attack* pada image fingerprinting karena semua salinan fingerprint harus tampak identik. Sedangkan pada software watermarking, untuk menghadapi *collusive attack* dapat dilakukan dengan memberikan transformasi obfuscation yang berbeda pada setiap salinannya, sehingga setiap salinan pasti mempunyai bentuk yang berbeda.

**4.1 Skema Pengujian Sistem**

Attack yang akan diuji-cobakan pada sistem adalah attack yang telah diujicobakan oleh peneliti-peneliti sebelumnya ditambah dengan kemungkinan penggunaan tool yang terbaru.

Attack utama yang diharapkan dapat terlewati adalah *dynamic analisis attack*, dimana berhasil tidaknya attack jenis ini sangat ditentukan oleh kemampuan attacker dalam mengamati perubahan-perubahan yang terjadi saat program dijalankan. Dengan demikian parameter untuk pengujian terhadap attack jenis ini adalah kabur tidaknya letak watermark (dalam hal ini dummy method). Diasumsikan bahwa letak dummy method adalah kabur jika kondisi-kondisi yang digunakan untuk memanggil dirinya bersifat dinamis (dapat bernilai *true* atau *false*).

Untuk pengujian terhadap static analisis dibatasi dengan penggunaan tool otomatis yang sudah ada dan dibedakan berdasarkan jenis attack-nya yaitu :

I. Obfuscator Attack

Obfuscator adalah suatu tool untuk merubah perangkat lunak kedalam bentuk lain tanpa merubah spesifikasi program. Proses transformasi kode ini kemungkinan dapat merubah kode watermark.

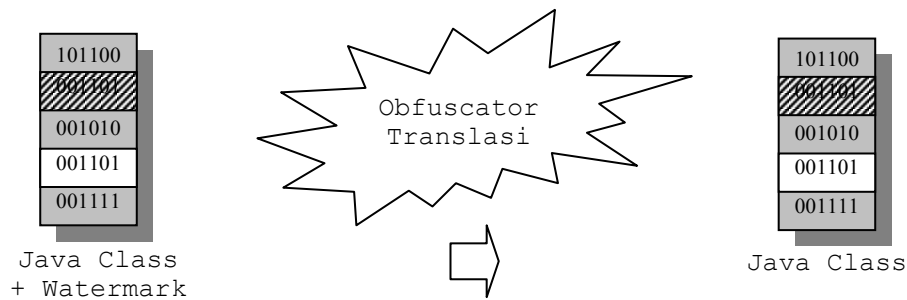

*Gambar 4.1 Jenis attack dengan menggunakan obfuscator tool*

Pada ujicoba ini digunakan tool *Zelix KlassMaster*, dimana tool ini mempunyai option yang lengkap untuk operasi obfuscation, seperti :

- *Obfuscate Control Flow* : [none, light, normal, aggressive]
- *Encrypt String Literal* : [none, light, aggressive, flow obfuscate]
- *Keep inner class information*
- *Line numbers tables* : [delete, scramble, keep]

II. Decompile-Recompile Attack

Attack ini dilakukan dengan men-dekompilasi perangkat lunak kembali ke kode sumber dengan menggunakan tool-tool yang sudah ada. Kode sumber yang dihasilkan selanjutnya dikompilasi kembali menjadi class file.

Attack jenis ini dapat merubah susunan kode watermark yang berarti membuat watermark menjadi rusak.

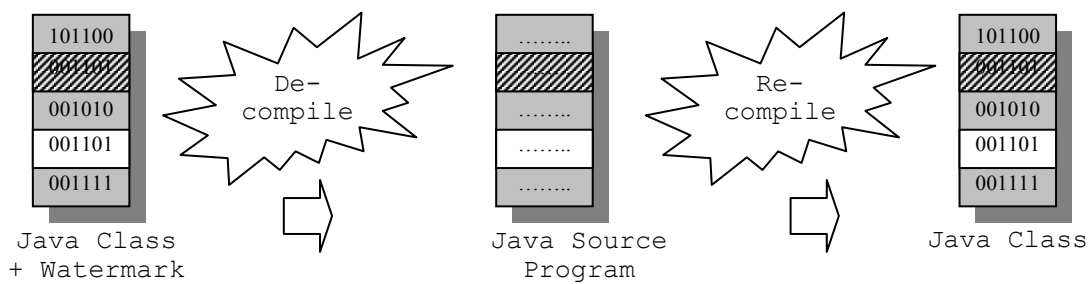

*Gambar 4.2 Jenis attack dengan menggunakan skema decompile-recompile*

Pada percobaan ini digunakan Java decompiler *Jad* dan *Jode* untuk menghasilkan kembali kode sumber. Kode sumber yang dihasilkan selanjutnya dikompilasi kembali dengan *Javac* (compiler asli dari JDK 1.4.0).

III. Trimmer Attack

Trimmer adalah suatu tool yang digunakan untuk membuang class, field dan method yang tidak digunakan dalam perangkat lunak tersebut.

Hal ini sangat menarik karena dummy method merupakan method yang tidak akan pernah di-eksekusi.

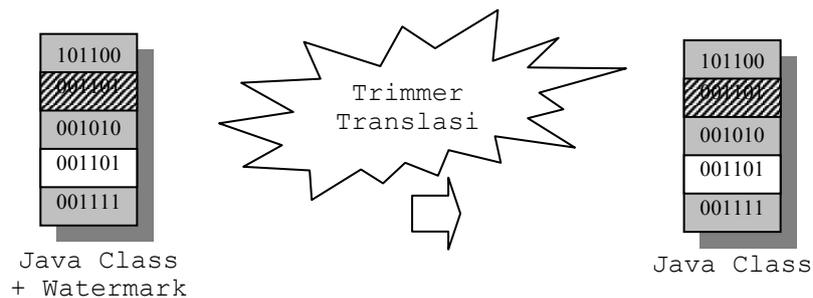

*Gambar 4.3 Jenis attack dengan menggunakan trimmer tool*

Pada percobaan ini digunakan *Zelix KlassMaster* yang memungkinkan user untuk menentukan aturan trim-nya. Parameter yang dibutuhkan adalah menentukan *Application type* dan *Application entry point*.

Contoh aturan trimmer :

```
Application type : Self contained application or applet
Application entry point class : MyClass

Exclude all methods:
   1) that are "public static "
   2) that have a name matching "main"
   3) that have arguments matching "java.lang.String[]"
   4) that are contained within a class:
        a) that is not in a package
        b) with a name that matches "MyClass"
Also exclude any class containing a matching method
```

IV. Dynamic Analysis Attack

Attack yang telah disebutkan sebelumnya adalah bersifat static analysis yaitu attack yang diberikan pada saat perangkat lunak tidak sementara dijalankan.

Dynamic analysis dilakukan untuk memantau jalannya perangkat lunak baik sifat atau perubahan sifat yang terjadi selama program berjalan. Hal ini penting karena seorang *expert attacker* yang dapat memantau heap, register selama eksekusi akan dapat me-lokalisir kode watermark (dalam hal ini dummy method) [PAL-2000].

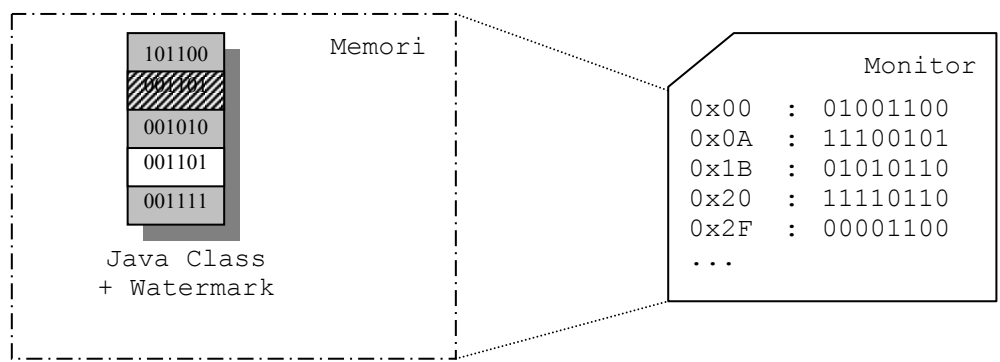

*Gambar 4.4 Jenis attack dengan menggunakan skema dynamic analysis*

Pada percobaan ini pengamatan dilakukan secara manual terhadap kondisi-kondisi boolean dalam program (baik kondisi boolean asli maupun opaque predicate) termasuk opaque predicate yang menjadi anggota kelompok yang menghasilkan dynamically opaque predicate.

## 4.2 Prosedur Pengujian Sistem

Langkah I : Mempersiapkan kode-kode sumber.

Kode-kode sumber dipersiapkan yang sebagian besar diambil dari program demo yang disertakan pada distribusi JDK 1.4.0.

Langkah II : Menyisipkan Dummy Method

Dummy method dipersiapkan 5 buah yang disisipkan secara acak pada program-program tersebut. Jumlah dummy method untuk setiap program bervariasi tergantung besar-kecilnya kode program.

Langkah III : Menyisipkan Dynamically Opaque Predicates

Membuat opaque-opaque predicates pada masing-masing program. Letak maupun jumlahnya tergantung besar-kecilnya program.

Langkah IV : Encoding Watermark

Pada semua dummy method disisipkan kode "ITS SURABAYA" sebagai kode watermark. Proses ini dilakukan secara otomatis dengan menggunakan tool yang telah dibuat.

Langkah V : Memberikan Attack

Untuk semua program diberikan attack-attack yang telah dipersiapkan.

Langkah VI : Decoding Watermark

Semua kode watermark lalu di-decode kembali, yang diharapkan seharusnya tetap tidak terganggu. Untuk dynamic analysis attack, yang diperhatikan adalah kemungkinan-kemungkinan yang dapat menyebabkan kode watermark dilokalisir.

## 4.3 Pengujian Sistem

Skenario Uji Coba I :

Terdapat aplikasi *Stylepad* (pengolah kata dalam bahasa Java) ingin di-watermark dengan kode rahasia "ITS SURABAYA".

Aplikasi ini mempunyai 5 buah file kode sumber :

| File | Ukuran Source | Ukuran Class |
|------|---------------|--------------|
| Stylepad.java | 7.997 bytes | 9.109 bytes |
| ElementTreePanel.Java | 16.799 bytes | 11.406 bytes |
| Notepad.java | 18.690 bytes | 26.921 bytes |
| HelloWorld.java | 5.139 bytes | 5.302 bytes |
| Wonderland.java | 8.329 bytes | 7.939 bytes |

Tahapan Off-line :

o   Menambahkan Dummy Method *Z* pada file *Stylepad.java*

```java
private void Z(int k){
    int i, j, tmp;
    int[] A;
    A = new int[100];
    A[0] = 5;
    A[1] = 7;
    A[2] = 1;
    A[3] = 6;
    A[4] = 4;
    System.out.println("k = " + k);
    for(i = 0; i < 5 ; i++){
        System.out.println("A["+ i + "] = " + A[i]);
    }
    for(i = 0; i < 4; i++){
        for(j = 1; j < 5; j++){
            if(A[j] < A[i]){
                tmp = A[j];
                A[i] = A[j];
                A[j] = tmp;
            }
        }
    }
    for(i = 0; i < 5 ; i++){
        System.out.println("A["+ i + "] = " + A[i]);
    }
    for(i = 0; i < 5 ; i++){
        for(j = 0; i < 100 ; j++){
            A[i] += k + j * 5;
        }
        System.out.println("A["+ i + "] = " + A[i]);
    }
}
```

- Membuat stuktur data dinamis (*Node.java*).

  ```
  public class Node {
      public boolean token;
      public Node head, tail;
      public Node() {
          this.token = false;
          this.head = this.tail = this;
      }
      public Node addNode() {
          Node p = new Node();
          p.head = this.tail;
          this.head = p;
          return p;
      }
      Node MoveNext () {
          return this.tail.head;
      }
      Node MoveBack () {
          return this.head.tail;
      }
  }
  ```

- Menambahkan dua buah node pada *Stylepad.java*

  ```
  public Stylepad() {
      ...
      g = new Node();
      g.token = true;
      h = new Node();
      h.token = true;
      p = g.addNode();
      q = h.addNode();
      ...
  }
  ```

- Menambahkan dua buah Thread *S* dan *T*

  ```
  ...
  t = new Thread(this);
  s = new Thread(this);
  t.start();
  s.start();
  ...
  ```

  dimana, kedua thread ini akan meng-update struktur data *G* dan *H*

```
...
public void run() {
    while (true) {
        Thread ct = Thread.currentThread();
        if (ct == t) {
            p = p.MoveNext();
        } else if (ct == s) {
            q = q.MoveBack();
        }
        try {
            if (ct == t) {
                Thread.sleep(12000);
            } else if (ct == s) {
                Thread.sleep(4000);
            }
        } catch (InterruptedException ie) {}
    }
}
...
```

- Membuat opaque predicates, termasuk yang menghubungkan dummy method dengan method lainnya

```
...
public void actionPerformed(ActionEvent e) {
    ...
    b2 = (h.equals(p) || p.token);
}
...
if (b2 && b1 && (g.equals(h))) {
    Z(10);
} else
    w.loadDocument();
...
```

Proses off-line ini merubah spesifikasi *Stylepad.java* menjadi :

| File | Ukuran Source | Ukuran Class |
|---|---|---|
| Stylepad.java | 9.551 bytes | 13.405 bytes |

Proses ini juga merubah waktu loading aplikasi ini :

| Aplikasi Stylepad | Loading Time (milidetik) | | | | |
|---|---|---|---|---|---|
| | I | II | III | IV | Rata-Rata |
| Sebelum | 10160 | 9720 | 9610 | 9670 | 9790 |
| Sesudah | 12300 | 9720 | 9880 | 9720 | 10405 |
| **Perubahan** | | | | | **615** |

Proses penambahan dummy method dan opaque predicates ini lalu diulangi untuk semua kode sumber dalam aplikasi sehingga diperoleh spesifikasi baru :

| File | Ukuran Source | Ukuran Class |
|---|---|---|
| Stylepad.java | 9.551 bytes | 13.405 bytes |
| ElementTreePanel.Java | 18.135 bytes | 13.158 bytes |
| Notepad.java | 20.009 bytes | 29.080 bytes |
| HelloWorld.java | 6.465 bytes | 6.839 bytes |
| Wonderland.java | 9.682 bytes | 9.476 bytes |

dengan perbandingan loading time :

| Aplikasi Stylepad | Loading Time (milidetik) | | | | |
|---|---|---|---|---|---|
| | I | II | III | IV | Rata-Rata |
| Sebelum | 10160 | 9720 | 9610 | 9670 | 9790 |
| Sesudah | 48440 | 37900 | 38120 | 36860 | 40330 |
| **Perubahan** | | | | | **30540** |

Untuk setiap penambahan dummy method dan dynamically opaque predicates terjadi perubahan waktu loading aplikasi sekitar `30540 / 5 = 6108` milidetik

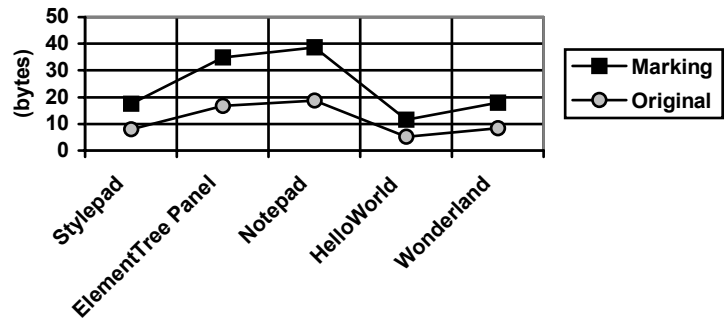

*Grafik 4.1 Perbandingan besar kode sumber aplikasi Stylepad sebelum dan sesudah dilakukan watermarking*

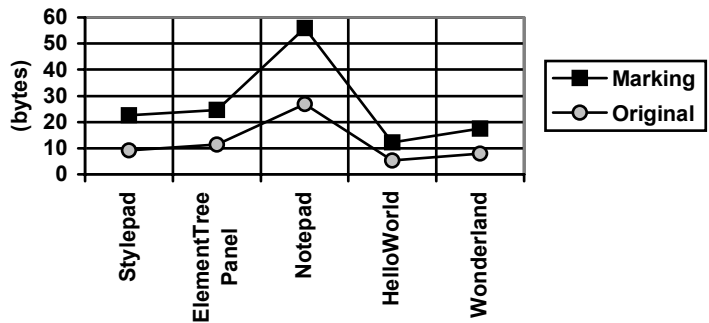

*Grafik 4.2 Perbandingan besar class-file aplikasi Stylepad
sebelum dan sesudah dilakukan watermarking*

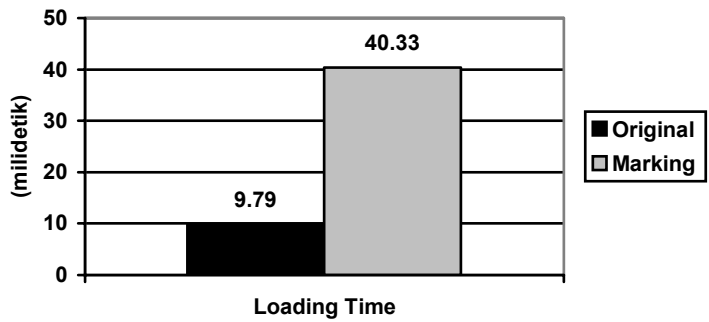

*Grafik 4.3 Perbandingan loading-time aplikasi Stylepad
sebelum dan sesudah dilakukan watermarking*

Tahapan On-line :

o   Masing-masing dummy method di-encode-kan kode "ITS SURABAYA"

| File | Jenis | Kunci |
|---|---|---|
| Stylepad.class | Normal | Yes |
| ElementTreePanel.class | Medium | No |
| Notepad.class | Medium | Yes |
| HelloWorld.class | Hard | No |
| Wonderland.class | Hard | Yes |

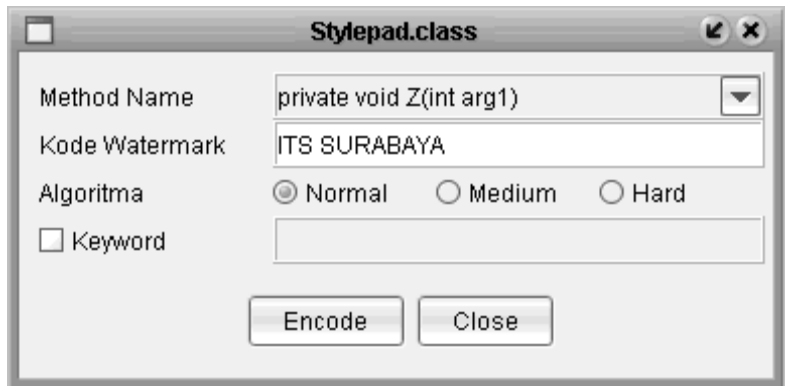

*Gambar 4.5 Contoh proses watermark encoding*

- o Setelah kode di-encode-kan pada semua dummy method, selanjutnya dilakukan attack pada aplikasi. Kode watermark dari aplikasi yang sudah dikenai attack tersebut lalu dicoba di-decode kembali.

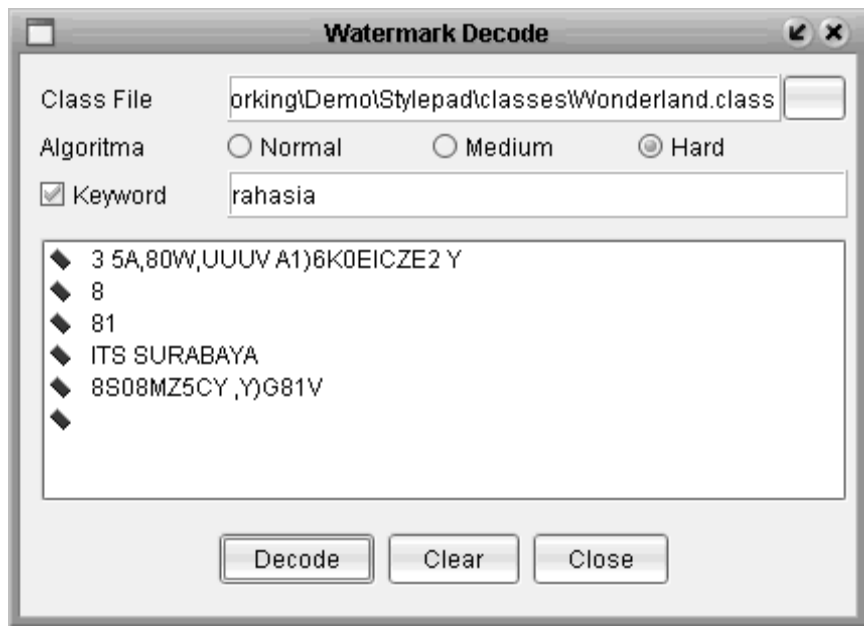

*Gambar 4.6 Contoh proses watermark de-coding*

Diperoleh hasil :

| File | Attack | | | |
|---|---|---|---|---|
| | Obfuscation Attack | Decompile-Recompile | Trimmer | Dynamic Analysis |
| Stylepad.class | Tidak | Ya | Ya | Dinamis |
| ElementTree Panel.class | Tidak | Ya | Ya | Dinamis |
| Notepad.class | Tidak | Ya | Ya | Dinamis |
| HelloWorld.class | Ya | Ya | Ya | Dinamis |
| Wonderland.class | Ya | Ya | Ya | Dinamis |
| Keterangan : <br> - **Tidak** berarti program tidak lolos melalui attack. <br> - **Ya** berarti lolos melalui attack. <br> - **Dinamis** berarti opaque predicate bersifat dinamis, kadang *true* atau *false*. | | | | |

Pengamatan *dynamically opaque predicates* :

```
P token = false, Q token = false, P == Q false
P token = false, Q token = true,  P == Q false
P token = false, Q token = false, P == Q false
P token = true,  Q token = false, P == Q false
P token = true,  Q token = true,  P == Q false
P token = true,  Q token = false, P == Q false
P token = false, Q token = false, P == Q false
P token = false, Q token = true,  P == Q false
P token = false, Q token = false, P == Q false
P token = true,  Q token = false, P == Q false
P token = true,  Q token = true,  P == Q false
P token = true,  Q token = false, P == Q false
P token = false, Q token = false, P == Q false
P token = false, Q token = true,  P == Q false
P token = false, Q token = false, P == Q false
P token = true,  Q token = false, P == Q false
P token = true,  Q token = true,  P == Q false
P token = true,  Q token = false, P == Q false
P token = false, Q token = false, P == Q false
P token = false, Q token = true,  P == Q false
P token = false, Q token = false, P == Q false
P token = true,  Q token = false, P == Q false
P token = true,  Q token = true,  P == Q false
P token = true,  Q token = false, P == Q false
```

Hasil pengamatan dynamically opaque predicates secara lengkap dapat dilihat pada lampiran II.

Tahapan-tahapan diatas selanjutnya diterapkan pada aplikasi lain.

Skenario Uji Coba II :

Aplikasi *FireWorks*, dengan file kode sumber :

| File | Ukuran Source | Ukuran Class |
|---|---|---|
| FireWorks.java | 923 bytes | 1.519 bytes |
| FCanvas.java | 461 bytes | 589 bytes |
| FireBall.java | 5.615 bytes | 3.800 bytes |
| Fragment.java | 2.783 bytes | 2.352 bytes |
| Launcher.java | 1.606 bytes | 1.385 bytes |

Setelah menambahkan watermark :

| File | Ukuran Source | Ukuran Class |
|---|---|---|
| FireWorks.java | 2.241 bytes | 2.875 bytes |
| FCanvas.java | 1.784 bytes | 2.236 bytes |
| FireBall.java | 6.867 bytes | 5.290 bytes |
| Fragment.java | 4.110 bytes | 3.991 bytes |
| Launcher.java | 2.859 bytes | 2.861 bytes |

| Aplikasi FireWorks | Loading Time (milidetik) | | | | |
|---|---|---|---|---|---|
| | I | II | III | IV | Rata-Rata |
| Sebelum | 2420 | 2360 | 2420 | 2300 | 2375 |
| Sesudah | 2470 | 2370 | 2360 | 2360 | 2390 |
| **Perubahan** | | | | | 15 |

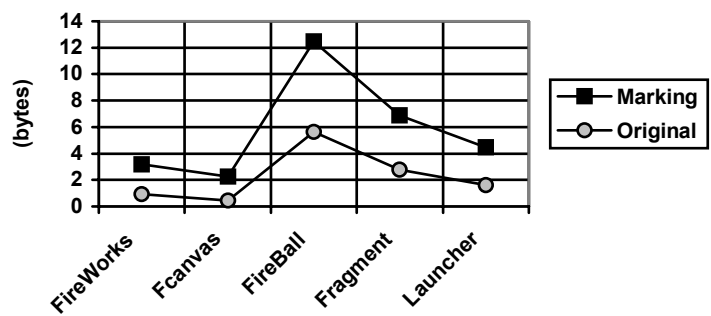

*Grafik 4.4 Perbandingan besar kode sumber aplikasi FireWorks sebelum dan sesudah dilakukan watermarking*

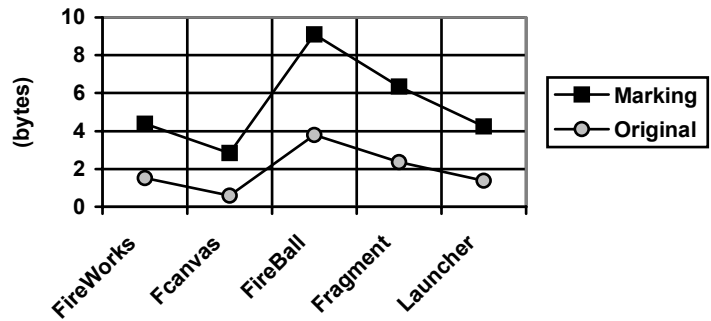

*Grafik 4.5 Perbandingan besar class-file aplikasi FireWorks
sebelum dan sesudah dilakukan watermarking*

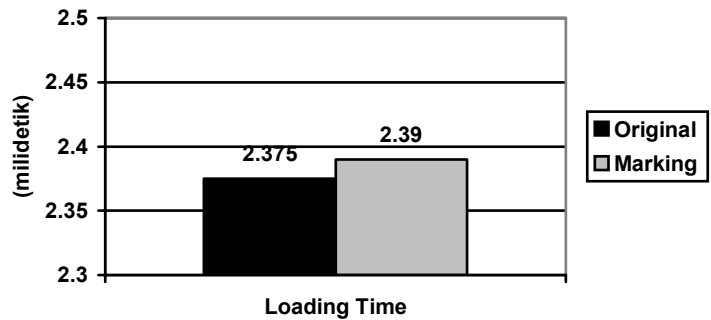

*Grafik 4.6 Perbandingan loading-time aplikasi FireWorks
sebelum dan sesudah dilakukan watermarking*

| File | Attack | | | |
|---|---|---|---|---|
| | Obfuscation Attack | Decompile-Recompile | Trimmer | Dynamic Analysis |
| FireWorks.class | Tidak | Ya | Ya | Dinamis |
| Fcanvas.class | Tidak | Ya | Ya | Dinamis |
| FireBall.class | Tidak | Ya | Ya | Dinamis |
| Fragment.class | Ya | Ya | Ya | Dinamis |
| Launcher.class | Ya | Ya | Ya | Dinamis |
| Keterangan : <br> - **Tidak** berarti program tidak lolos melalui attack. <br> - **Ya** berarti lolos melalui attack. <br> - **Dinamis** berarti opaque predicate bersifat dinamis, kadang *true* atau *false*. | | | | |

Pemberian attack selengkapnya dapat dilihat pada lampiran III.

**4.4 Analisa Sistem**

Analisa ini dimaksudkan untuk menjawab pertanyaan yang menjadi acuan dalam pengujian sistem sebagaimana yang dijelaskan pada awal bab ini sesuai dengan hasil yang diperoleh dari pengujian.

**(i) Pada struktur data apa sebaiknya watermark diletakkan.**

Pada dummy method watermarking, kode watermark diletakkan pada suatu method (dummy). Untuk memperoleh dummy method yang cukup besar, sebaiknya method ini berisi struktur data operasi aritmatik.

Hal ini erat hubungannya dengan salah satu karakteristik software watermarking *high data rate*, tempat watermark harus cukup besar menampung kode karena apabila kode terpotong (karena dummy method terlalu kecil) dapat menyebabkan klaim kepemilikan bisa tidak berarti.

**(ii) Bagaimana kita meng-ekstrak watermark dan meng-klaim bahwa itu adalah milik kita.**

Hal ini erat kaitannya dengan pertanyaan pertama tadi, bahwa untuk dapat meng-klaim bahwa itu adalah milik kita, tentu saja kode yang di-ekstrak harus sempurna.

Untuk meng-ekstrak watermark dilakukan dengan melakukan proses dekoding pada semua method dalam setiap class file. Hal ini karena letak dummy method bisa berada dimana saja dalam setiap class file. Untuk proses ini dapat dibuatkan automatic tool yang akan melakukan proses penggantian op-code ke dalam bit-bit sesuai aturan pengkodean bit dan selanjutnya mengganti bit-bit menjadi urutan karakter sesuai dengan aturan pengkodean karakter.

**(iii) Bagaimana kita mencegah seorang attacker merusak watermark.**

Dari hasil uji coba diatas, ada kemungkinan watermark gagal melalui proses transformasi (distortive attack). Untuk itu proses injeksi watermark sebaiknya hanya menggunakan replacing-opcodes (walaupun ini tentu saja memerlukan dummy method yang lebih besar).

Untuk collusive attack dapat digunakan transformasi obfuscation yang berbeda-beda sehingga bentuk program berbeda satu dengan lainnya.

Untuk additive attack, tidak ada teknik software watermarking yang dapat 100% bertahan . Untuk mengatasi hal ini dapat dilakukan pencegahan supaya proses tersebut tidak dapat dilakukan pada aplikasi, misalnya melengkapi teknik ini dengan tamper-proofing yang akan merusak aplikasi apabila attacker berusaha memasukkan kode watermark lain (program mengalami perubahan).

**(iv) Seberapa besar watermark mempengaruhi performance program.**

Performance aplikasi akan berkurang berbanding lurus dengan besar dan jumlah dummy method serta opaque predicates yang ditambahkan dalam program. Semakin banyak dummy method, opaque predicates yang digunakan, maka dapat diharapkan tingkat resistansi yang lebih baik, tetapi menambah penggunaan sumber daya lebih, misal memory, media penyimpanan dll.

Untuk mengatasi hal ini, watermark mungkin dapat diterapkan pada class file yang berisi algoritma atau struktur data penting saja pada aplikasi.

# BAB V

# KESIMPULAN DAN SARAN

## 5.1 Kesimpulan

Setiap teknik software watermarking selalu memperhatikan hubungan yang terjadi antara *resilient*, *data rate*, *cost* dan *stealth*. Misalnya, *resilient* suatu teknik watermarking dapat dengan mudah ditingkatkan dengan melakukan *redundancy* (misal, mengulang-ulang kode watermark dalam aplikasi), akan tetapi hal ini akan membuat aplikasi membutuhkan lebih banyak sumber daya.

Dari uji coba diketahui bahwa teknik dummy method watermarking yang dikembangkan dengan dynamically opaque predicates ini, penambahan satu kode watermark menambah ukuran file sekitar 3.854 bytes untuk dummy method yang mampu menampung sekitar 15 karakter pada *replacing-opcodes* dengan penambahan dua struktur data dinamis, dua buah thread serta penambahan dua opaque predicates. Loading-time aplikasi bertambah rata-rata 6108 milidetik.

Pada teknik ini hal utama yang menjadi perhatian adalah dummy method harus tidak pernah dieksekusi. Hasil uji coba menunjukkan syarat tersebut dapat terpenuhi. Penerapan dynamically opaque predicates membuat kondisi untuk meng-eksekusi dummy method seolah-olah menjadi dinamis. Nilai-nilai opaque predicates untuk kondisi tersebut berubah secara acak dalam setiap run pengamatan.

Untuk attack terhadap software watermarking dapat disimpulkan :

o Distortive Attack.

Proses transformasi seperti transformasi obfuscation, kompilasi-dekompilasi dapat dihadapi dengan injeksi watermark menggunakan replacing-opcodes

- o  Collusive Attack.

    Menghadapi collusive attack dapat digunakan transformasi obfuscation yang berbeda-beda pada setiap salinan program sehingga bentuk program berbeda satu dengan lainnya.

- o  Additive Attack.

    Walaupun tidak ada teknik software watermarking yang dapat 100% bertahan menghadapi additive attack [COL-1999], hal tersebut dapat dicegah dengan memberikan tamper-proofing yang akan merusak aplikasi apabila attacker berusaha memasukkan kode watermark lain karena program mengalami perubahan.

## 5.2 Saran

Kekurangan utama skema ini adalah pada adanya tambahan pekerjaan manual pada tahapan persiapan watermarking. Penelitian ini mungkin dapat dilanjutkan untuk meng-otomatisasi hal tersebut.

Hal lain juga dimungkinkan untuk melengkapi teknik ini dengan menambahkan tamper-proofing untuk mencegah attacker berusaha menghilangkan dummy method yang merupakan tempat kode watermark atau menambahkan kode watermark sendiri. Penggunaan skema watermarking yang berbeda dan berlapis untuk melindungi suatu aplikasi tentu akan lebih baik. Untuk itu dalam melakukan watermarking sebaiknya tidak satu skema watermarking saja.

# DAFTAR PUSTAKA

## Lampiran I. Java Class File Format

Keterangan :
    Huruf u menandakan *unsigned-byte* yang diikuti angka sebagai jumlah byte.

```
ClassFile {
    u4   magic;
    u2   minor_version;
    u2   major_version;
    u2   constant_pool_count;
    cp_info constant_pool [constant_pool_count-1];
    u2   access_flags;
    u2   this_class;
    u2   super_class;
    u2   interfaces_count;
    u2   interfaces [interfaces_count];
    u2   fields_count;
    field_info fields[fields_count];
    u2   methods_count;
    method_info methods [methods_count];
    u2   attributes_count;
    attribute_info attributes [attributes_count];
}

cp_info {
    u1   tag;
    u1   info[];
}

tag
   CONSTANT_Class              7
   CONSTANT_Fieldref           9
   CONSTANT_Methodref          10
   CONSTANT_InterfaceMethodref 11
   CONSTANT_String             8
   CONSTANT_Integer            3
   CONSTANT_Float              4
   CONSTANT_Long               5
   CONSTANT_Double             6
   CONSTANT_NameAndType        12
   CONSTANT_Utf8               1

CONSTANT_Class_info {
    u1   tag;
    u2   name_index;
}

CONSTANT_Fieldref_info {
    u1   tag;
    u2   class_index;
    u2   name_and_type_index;
}
```

```
CONSTANT_Methodref_info {
    u1  tag;
    u2  class_index;
    u2  name_and_type_index;
}

CONSTANT_InterfaceMethodref_info {
    u1  tag;
    u2  class_index;
    u2  name_and_type_index;
}

CONSTANT_String_info {
    u1  tag;
    u2  string_index;
}

CONSTANT_Integer_info {
    u1  tag;
    u4  bytes;
}

CONSTANT_Float_info {
    u1  tag;
    u4  bytes;
}

CONSTANT_Long_info {
    u1  tag;
    u4  high_bytes;
    u4  low_bytes;
}

CONSTANT_Double_info {
    u1  tag;
    u4  high_bytes;
    u4  low_bytes;
}

CONSTANT_NameAndType_info {
    u1  tag;
    u2  name_index;
    u2  descriptor_index;
}

CONSTANT_Utf8_info {
    u1  tag;
    u2  length;
    u1  bytes[length];
}

access_flag
    ACC_PUBLIC      0x0001
    ACC_FINAL       0x0010
    ACC_SUPER       0x0020
    ACC_INTERFACE   0x0200
    ACC_ABSTRACT    0x0400
```

```
field_info {
    u2   access_flags;
    u2   name_index;
    u2   descriptor_index;
    u2   attributes_count;
    attribute_info attributes [attributes_count];
}

access_flag
   ACC_PUBLIC       0x0001
   ACC_PRIVATE      0x0002
   ACC_PROTECTED    0x0004
   ACC_STATIC       0x0008
   ACC_FINAL        0x0010
   ACC_VOLATILE     0x0040
   ACC_TRANSIENT    0x0080

method_info {
    u2   access_flags;
    u2   name_index;
    u2   descriptor_index;
    u2   attributes_count;
    attribute_info attributes [attributes_count];
}

access_flag
   ACC_PUBLIC       0x0001
   ACC_PRIVATE      0x0002
   ACC_PROTECTED    0x0004
   ACC_STATIC       0x0008
   ACC_FINAL        0x0010
   ACC_SYNCHRONIZED 0x0020
   ACC_NATIVE       0x0100
   ACC_ABSTRACT     0x0400
   ACC_STRICT       0x0800

attribute_info {
    u2   attribute_name_index;
    u4   attribute_length;
    u1   info[attribute_length];
}

ConstantValue_attribute {
    u2   attribute_name_index;
    u4   attribute_length;
    u2   constantvalue_index;
}

Code_attribute {
    u2   attribute_name_index;
    u4   attribute_length;
    u2   max_stack;
    u2   max_locals;
    u4   code_length;
    u1   code[code_length];
    u2   exception_table_length;
    {   u2   start_pc;
        u2   end_pc;
```

```
        u2    handler_pc;
        u2    catch_type;
    }   exception_table [exception_table_length];
    u2   attributes_count;
    attribute_info attributes [attributes_count];
}

Synthetic_attribute {
    u2   attribute_name_index;
    u4   attribute_length;
}

Exceptions_attribute {
    u2   attribute_name_index;
    u4   attribute_length;
    u2   number_of_exceptions;
    u2   exception_index_table [number_of_exceptions];
}

InnerClasses_attribute {
    u2   attribute_name_index;
    u4   attribute_length;
    u2   number_of_classes;
    {   u2 inner_class_info_index;
        u2 outer_class_info_index;
        u2 inner_name_index;
        u2 inner_class_access_flags;
    }   classes [number_of_classes];
}

inner_class_access_flags
    ACC_PUBLIC       0x0001
    ACC_PRIVATE      0x0002
    ACC_PROTECTED    0x0004
    ACC_STATIC       0x0008
    ACC_FINAL        0x0010
    ACC_INTERFACE    0x0200
    ACC_ABSTRACT     0x0400

SourceFile_attribute {
    u2   attribute_name_index;
    u4   attribute_length;
    u2   sourcefile_index;
}

LineNumberTable_attribute {
    u2   attribute_name_index;
    u4   attribute_length;
    u2   line_number_table_length;
    {   u2   start_pc;
        u2   line_number;
    } line_number_table [line_number_table_length];
}

LocalVariableTable_attribute {
    u2   attribute_name_index;
    u4   attribute_length;
    u2   localvariable_table_length;
```

```
        {   u2   start_pc;
            u2   length;
            u2   name_index;
            u2   descriptor_index;
            u2   index;
        } local_variable_table [localvariable_table_length];
}

Deprecated_attribute {
    u2   attribute_name_index;
    u4   attribute_length;
}
```

## Lampiran II. Pengamatan Dynamically Opaque Predicates

- MyClass

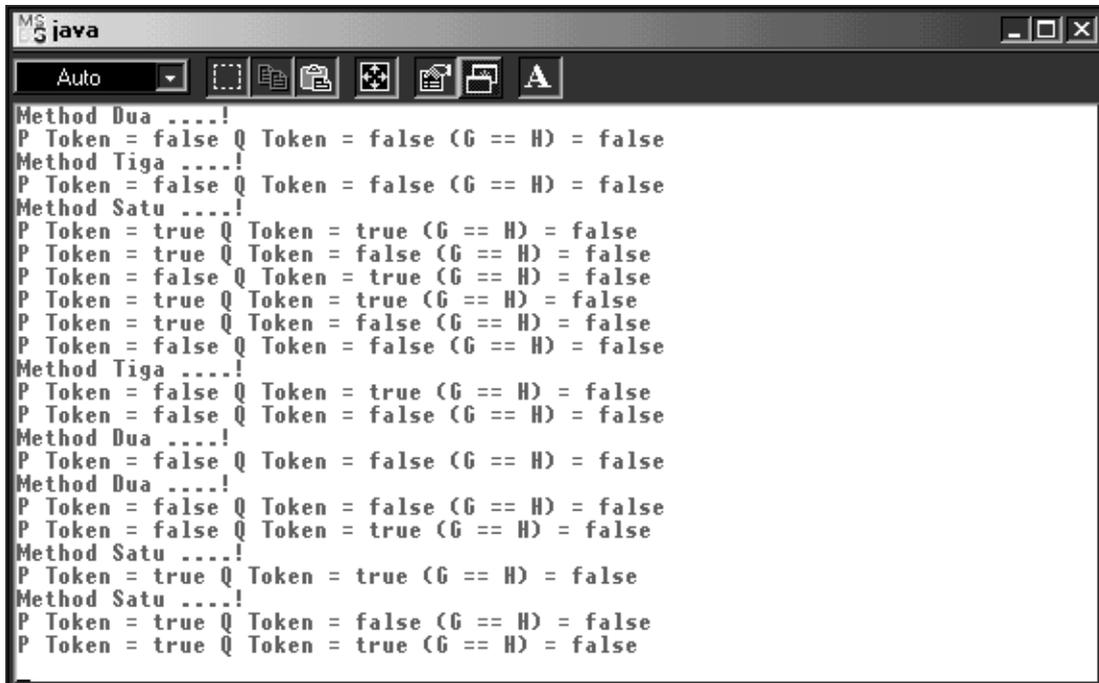

- Stylepad

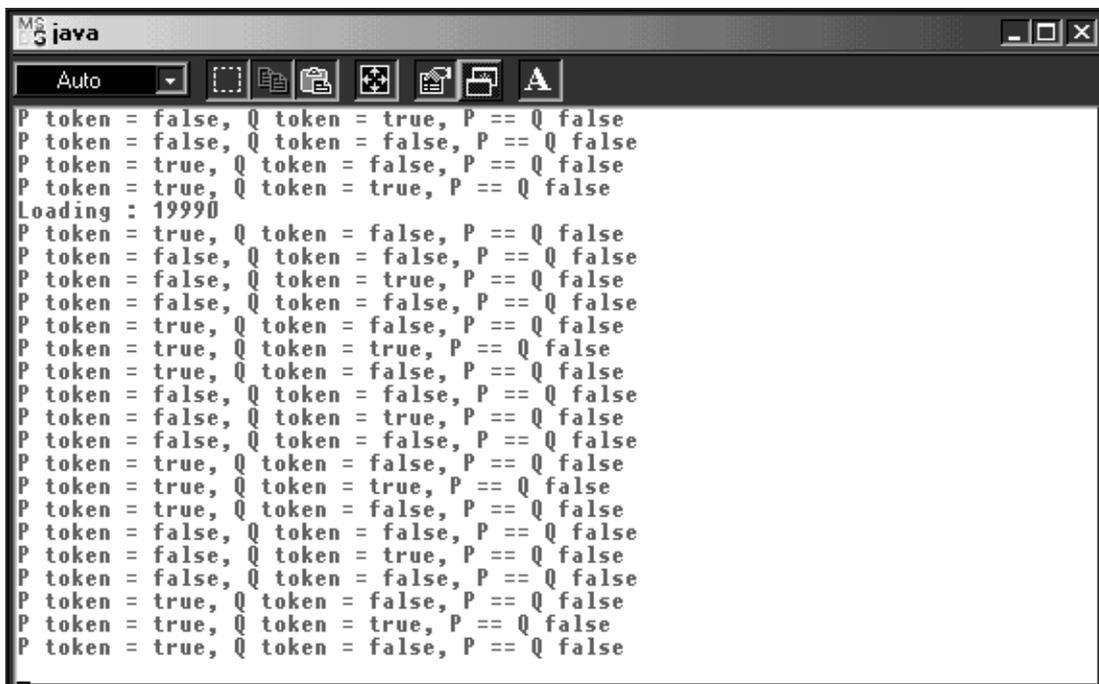

o  Modelling

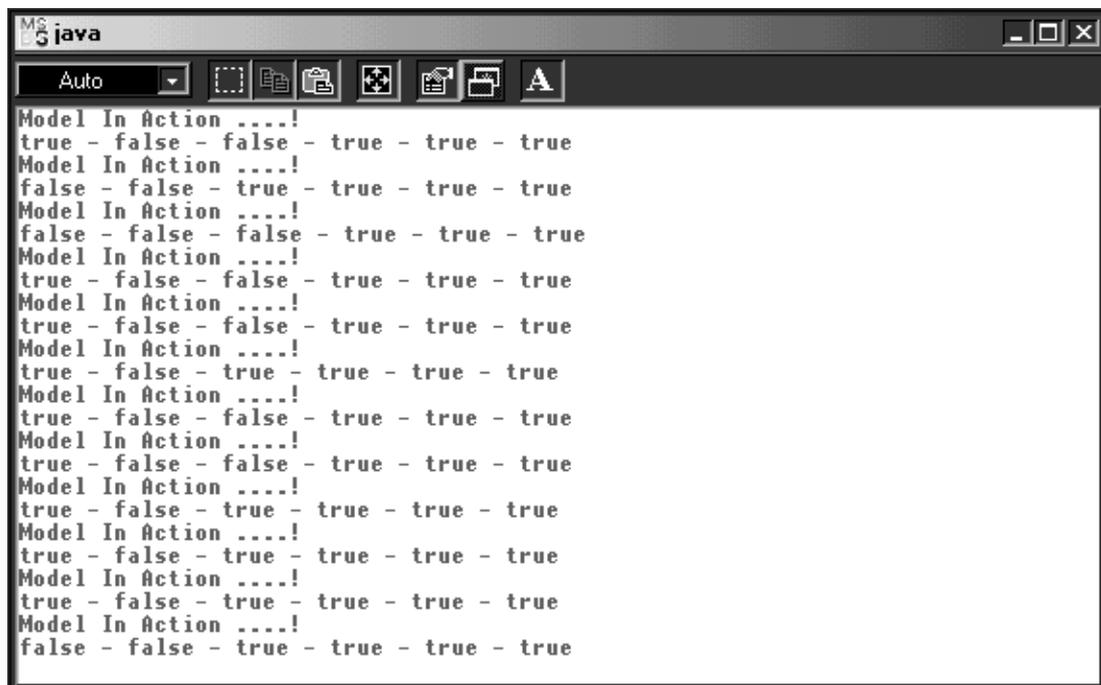

o  FireWorks

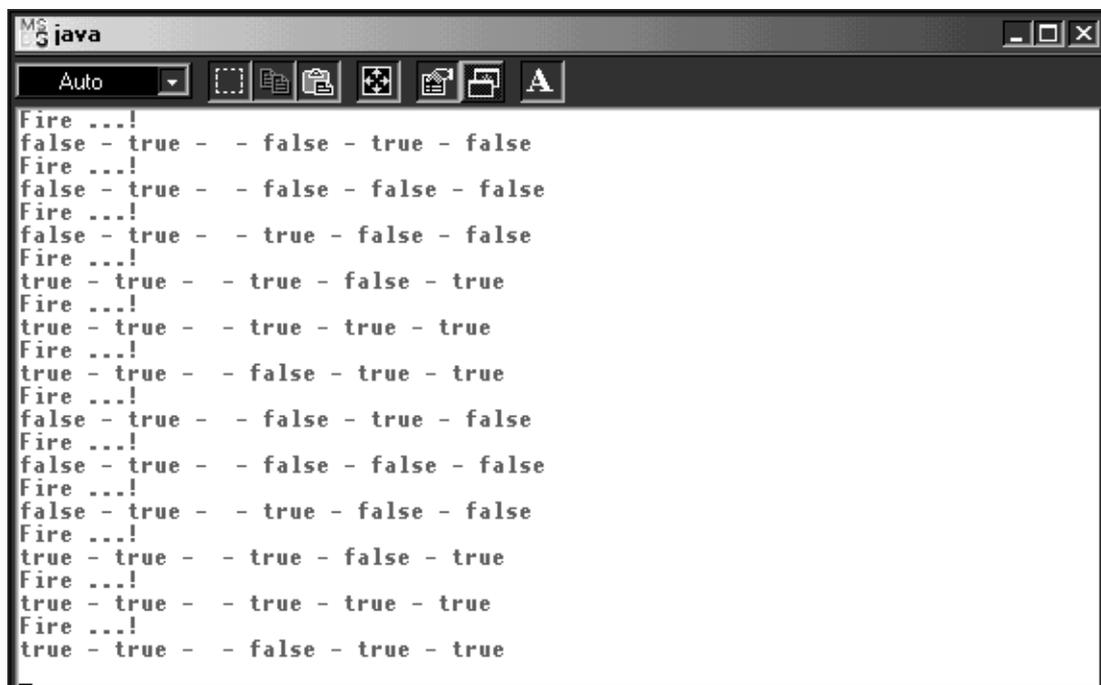

**Lampiran III. Pemberian Attack**

o   Obfuscation Attack

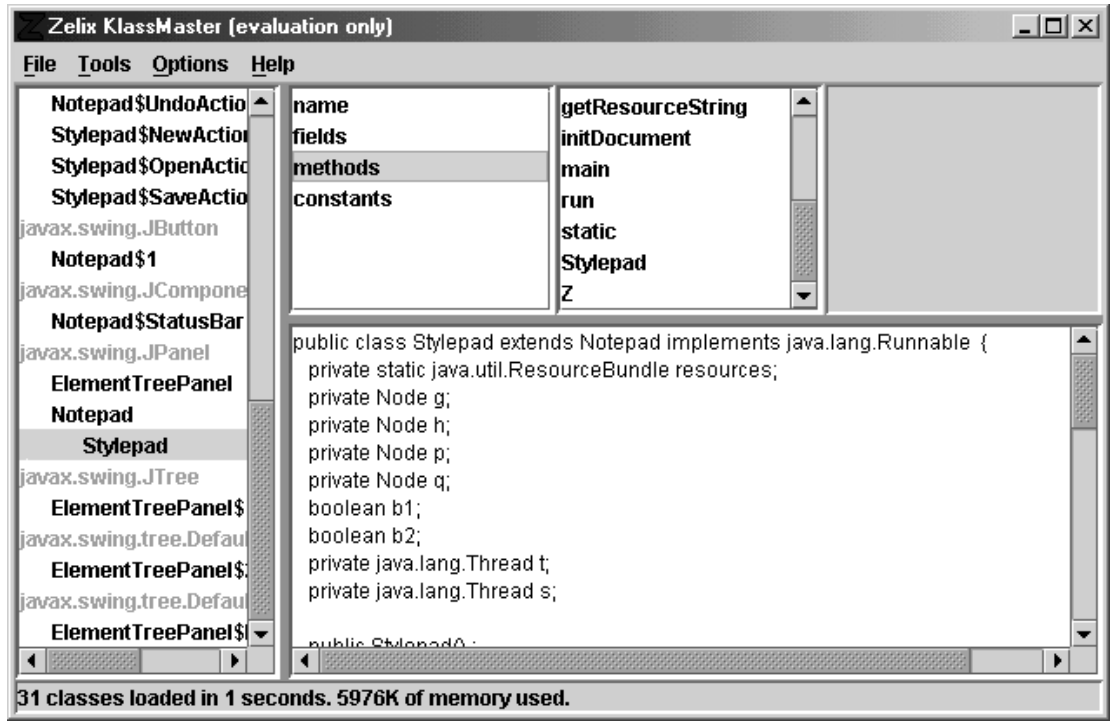

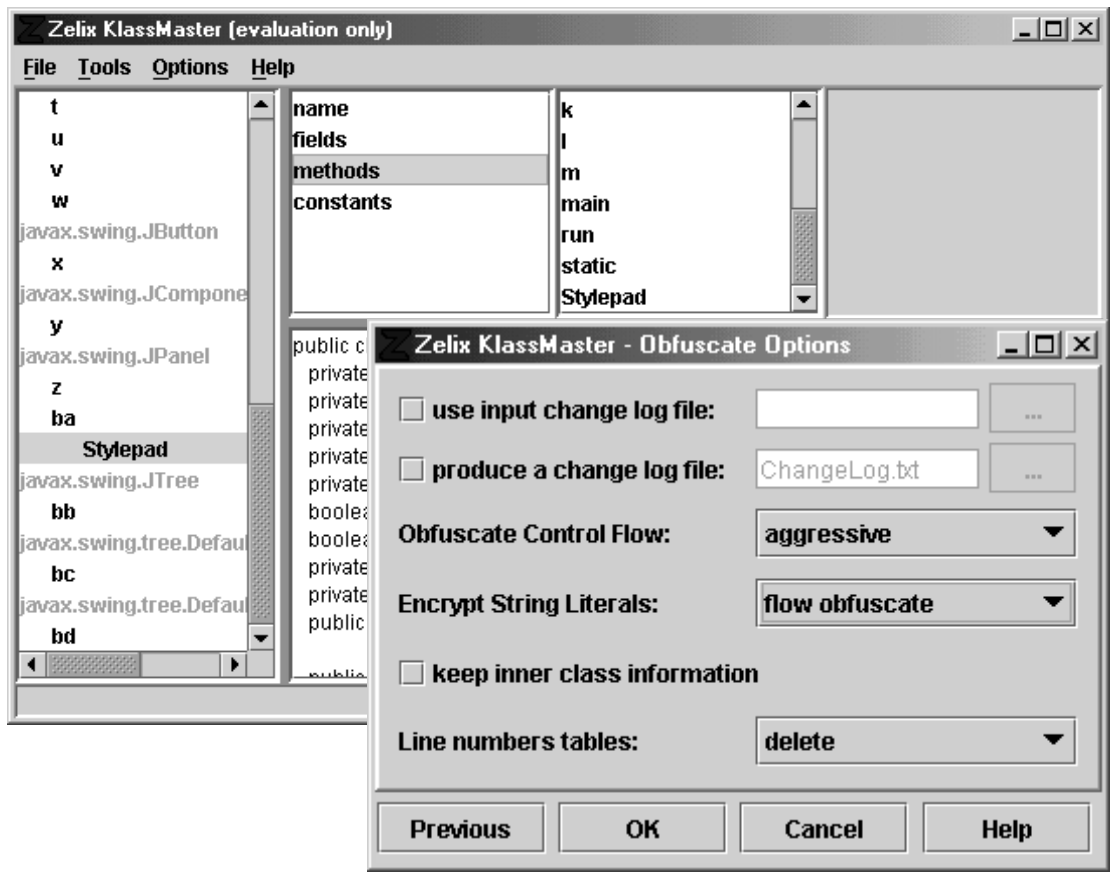

o   Decompile-Recompile Attack

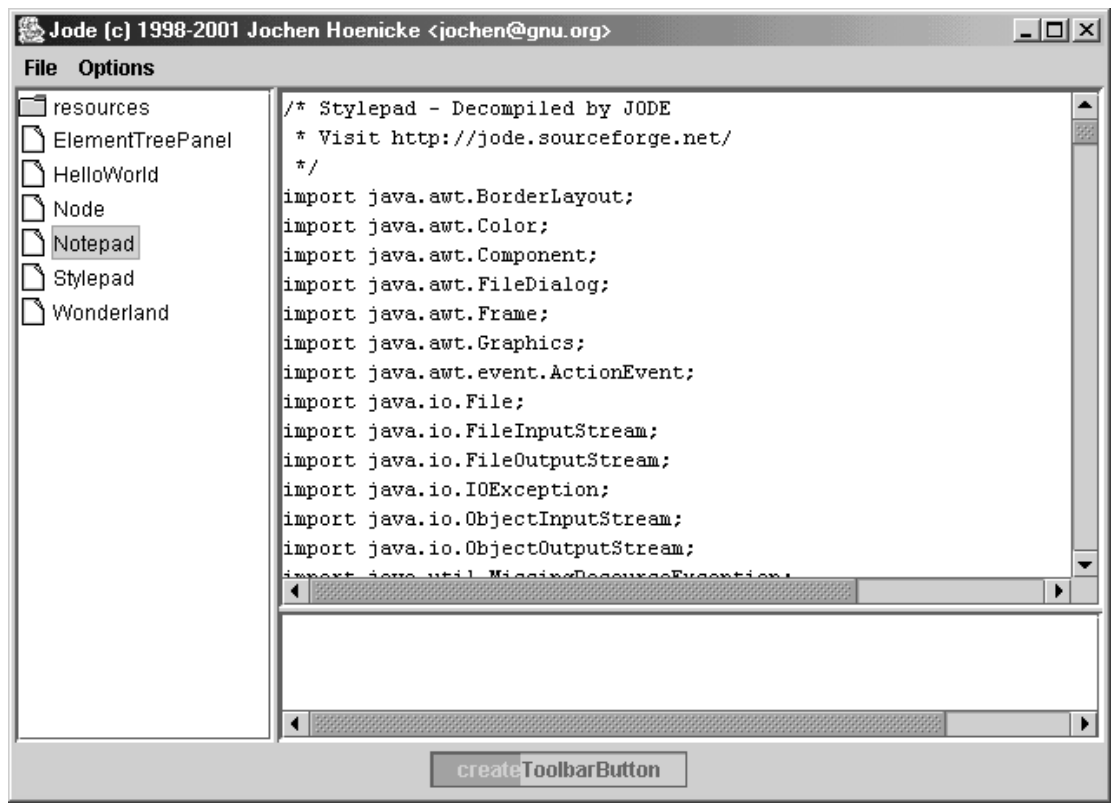

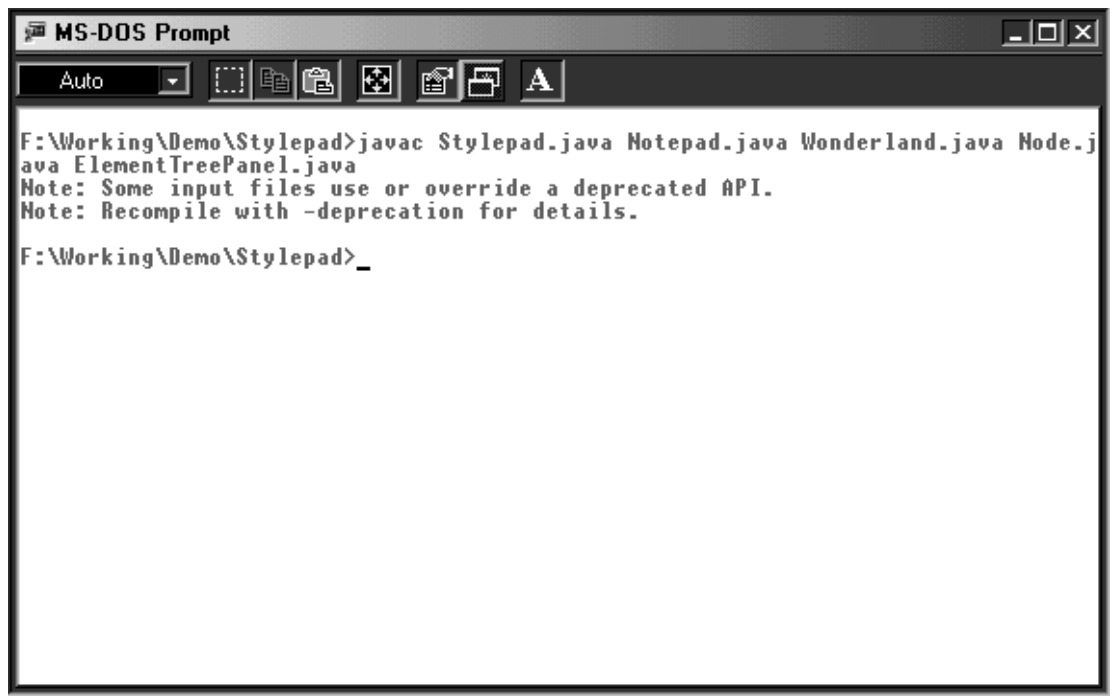

o   Trimmer Attack

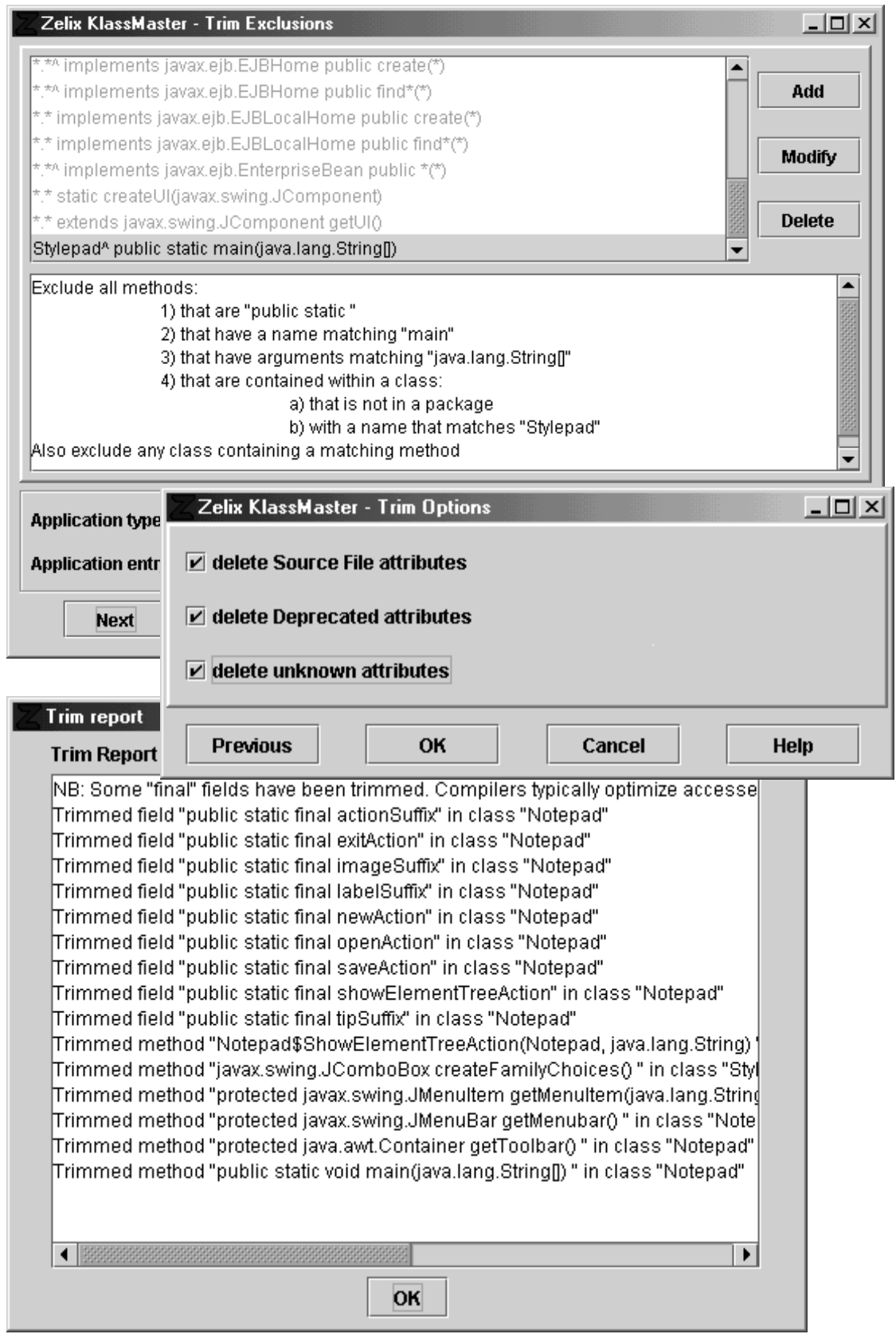

## Lampiran IV. Contoh Dummy Method

```java
private void R(int k){
    int i, j;
    for(i = 0; i < 100 ; i++){
        if(k % i != 0) {
            System.out.println("no." + i + " is OK.");
        }
    }
    for(i = 0; i < 10 ; i++){
        for(j = 0; j < 10 ; j++){
            k = k * 10 + i * 20 + j * 30;
        }
        for(j = 0; j < 50 ; j++){
            k+=j*3;
        }
    }
    System.out.println("k = " + k);
    for(i = 0; i < 20 ; i++){
        k+=i*5;
    }
    System.out.println("k = " + k);
}

private void S(int k){
    int i, j;
    int[] A;
    A = new int[100];
    A[0] = 105;
    A[1] = 127;
    A[2] = 51;
    A[3] = 16;
    A[4] = 44;
    A[5] = 74;
    A[6] = 84;
    System.out.println("k = " + k);
    for(i = 0; i < 7 ; i++){
        System.out.println("A["+ i + "] = " + A[i]);
    }
    for(i = 0; i < 7 ; i++){
        A[i] = k * i;
        System.out.println("A["+ i + "] = " + A[i]);
    }
    for(i = 0; i < 100 ; i++){
        A[i] += k + i * 5;
        System.out.println("A["+ i + "] = " + A[i]);
    }
}
```

```java
private void X(int k){
    int i, j;
    for(i = 0; i < 10 ; i++)
        for(j = 0; j < 10 ; j++) k+=i*10+j;
    System.out.println("k = " + k);
    for(i = 0; i < 20 ; i++)
        for(j = 0; j < 30 ; j++) k+=i*3-j;
    System.out.println("k = " + k);
    for(i = 0; i < 25 ; i++)
        for(j = 0; j < 20 ; j++) k+=i*4-j*3;
    System.out.println("k = " + k);
}

private void Y(int k){
    int i, j;
    int t;
    int tmp;
    int[] A;
    if(k > 100) return;
    A = new int[100];
    for(i = 0; i < 100; i++){
        A[i] = i * 10 + k;
    }
    t = 0;
    for(i = 0; i < k; i++){
        t += A[i]/A[i-k];
    }
    System.out.println("k = " + k);
    System.out.println("t = " + t);
    for(i = 0; i < 100 ; i++){
        for(j = 0; j < k ; j++){
            A[i] = k + j;
        }
        System.out.println("A[" + i + "] = " + A[i]);
    }
    for(i = 0; i < 100 ; i++)
        for(j = 0; j < 100 ; j++) k += i * 5;
    System.out.println("k = " + k);
}

private void Z(int k){
    int i, j;
    int tmp;
    int[] A;
    A = new int[100];
    A[0] = 5;
    A[1] = 7;
    A[2] = 1;
    A[3] = 6;
    A[4] = 4;
    System.out.println("k = " + k);
    for(i = 0; i < 5 ; i++){
        System.out.println("A["+ i + "] = " + A[i]);
    }
    for(i = 0; i < 4; i++){
        for(j = 1; j < 5; j++){
```

```
                if(A[j] < A[i]){
                    tmp = A[j];
                    A[i] = A[j];
                    A[j] = tmp;
                }
            }
        }
        for(i = 0; i < 5 ; i++){
            System.out.println("A["+ i + "] = " + A[i]);
        }
        for(i = 0; i < 5 ; i++){
            for(j = 0; i < 100 ; j++){
                A[i] += k + j * 5;
            }
            System.out.println("A["+ i + "] = " + A[i]);
        }
    }
}
```